# Coarse Grained modeling of Zeolitic Imidazolate Framework-8 using MARTINI Force Fields

Cecilia M. S. Alvares,[1] Guillaume Maurin,[1] Rocio Semino[2]*

[1] ICGM, Univ. Montpellier, CNRS, ENSCM, Montpellier, France
[2] Sorbonne Université, CNRS, Physico-chimie des Electrolytes et Nanosystèmes Interfaciaux, PHENIX, F-75005 Paris, France
* rocio.semino@sorbonne-universite.fr

*In this contribution, the well-known MARTINI particle-based coarse graining approach is tested for its ability to model the ZIF-8 metal-organic framework. Its capability to describe structure, lattice parameters, thermal expansion, elastic constants and amorphization is evaluated. Additionally, the less coarsened models were evaluated for reproducing the swing effect and the host-guest interaction energies were analyzed. We find that MARTINI force fields successfully capture the structure of the MOF for different degrees of coarsening, with the exception of the MARTINI 2.0 models for the less coarse mapping. MARTINI 2.0 models predict more accurate values of $C_{11}$ and $C_{12}$, while MARTINI 3.0 has a tendency to underestimate them. Amongst the possibilities tested, the choice of bead flavors within a particular MARTINI version appears to have a less critical impact in the simulated properties of the empty framework. None of the CG models investigated were able capture the amorphization nor the swing effect within the scope of MD simulations. A perspective on the importance of having a proper LJ parametrization for modeling guest-MOF and MOF-MOF interactions is highlighted.*

## INTRODUCTION

Metal-Organic Frameworks (MOFs) have been intensely investigated in the past few decades on account of their wide range of possible applications to processes that constitute important industrial and environmental challenges, including drug-delivery, catalysis and gas adsorption/separation among others.[1-5] MOFs are formed of metal ions or metallic oxo-clusters bound together *via* multidentate organic ligands. Their large variety in compositional space confers them high tunability of their permanent porosities, both in terms of chemistry and geometrical features. Computer simulation techniques have largely contributed to developing MOF research by providing information that is complementary to that obtained by current direct experimental techniques.[6-12] In face of the old-dated acknowledgment of the atomistic nature of matter, different formalisms have been considered - or even elaborated - to address the dynamics of the structural units (electrons, atoms), being classical and quantum mechanics the most traditional ones. While quantum mechanics, within the Born Oppenheimer approximation, evolves the system by treating the dynamics in the electronic level, classical mechanics is used, generally, to treat the dynamics of the atoms constituting the given system. Assessment of system dynamics may also be carried out at a macroscopic level upon considering reaction rates and statistical mechanics.

The choice of modeling resolution is driven by a mindful compromise between the kind of information that is desired as an output of the simulation and computational cost. While atom-level simulation methods, such as all-atom molecular dynamics, in general only permit treating systems sized of a few unit cells, techniques such as kinetic Monte Carlo, allow answering macroscopic-level questions more effectively, albeit losing molecular detail. Between these two, there is a gap in simulation resolution that corresponds to mesoscopic domains. Bridging this gap is crucial to help understand important phenomena such as collective pore breathing modes, gas transport within the hierarchical porosity formed in MOF-based composites and large-scale defects impact in local structure. Aiming to bridge the gap whilst still having computationally affordable simulations, the notion of creating particle-based coarse grained representations arose.[13-19] These are generated by lumping sets of atoms of the system into new entities, meant to be used as the new structural units of the system whose interactions are modeled classically. The new entities, referred to as superatoms or beads, have their types (or flavors) primarily defined by the chemical identity of the atoms that have been lumped and, possibly, by the identity of their neighbors as well. Ultimately, a coarse grained representation, also referred to as mapping, is somewhat subjective in the sense that there is no compulsory prescription on how to decide which atoms are going to be lumped into a bead.

Up to date, only one research contribution has been devoted to the development of particle-based coarse-grained (CG) force fields for modeling MOFs.[20] In their pioneering work, Dürholt and collaborators developed the first force field of this kind for the HKUST-1 MOF *via* a genetic algorithm optimization of parameters to reproduce the Hessian matrix of a reliable benchmark atomistic model. This CG parameterization algorithm could be classified as a force matching method, but other algorithms that focus on reproducing different aspects of benchmark models exist, including relative entropy and structural methods,[21] as well as more recent machine-learning based approaches.[22] Among the existing coarse graining approaches, the MARTINI force-fields have particularly emerged in the bioinformatics community.[23,24] They aim for transferability, reproducing thermodynamic properties for a large database of compounds and thus facilitating their implementation and use. To the best of our knowledge, these force-fields have not yet been tested for MOFs.

In this contribution, we test the MARTINI force fields for modeling the archetypal MOF ZIF-8.[25] This MOF belongs to the zeolitic imidazolate framework group, because its topology resembles an existing zeolite topology: *sodalite*. In ZIF-8, the $Zn^{2+}$ cations play the role of the silicon atoms, each of them is tetrahedrally coordinated to four 2-methylimidazolate ($mIm^-$) bidentate ligands instead of oxygens. ZIF-8 thus contains sodalite cages of 11.6 Å size accessible through small pore windows of 3.5 Å, which make it a highly promising candidate for alkane/alkene gas separation, among many other applications.[26,27] This MOF's associated high thermal and chemical stability as well as hydrophobic character add to its success. Even though many high quality atomistic force fields have been particularly developed to model ZIF-8,[28-32] there is no CG force field to model it up to date.

In this work, we systematically apply MARTINI 2.0 and 3.0 force fields to model ZIF-8 in the CG level for mappings of different coarsening degrees. Different bead flavor choices are also tested within each mapping. We assess the quality of the force fields by evaluating their performance in reproducing structure, lattice parameters, elastic constants, volume expansion coefficients and guest-induced swing effect. The best model for each mapping is selected, and results are explained in terms of the underlying physics.

This article is organized as follows. The methodological approach to the parameterization of force fields is described in Section II. Section III summarizes the critical assessment of the performance of the different models and the selection of the best model for each mapping. Finally, conclusions are given at Section IV.

# METHODOLOGY

All simulations within this work were performed *via* the LAMMPS open source molecular dynamics code (versions 7 Aug 2019 and 12 Dec 2018).[33]

## a) Classical models set-up

Force fields of a given form can be parameterized for reproducing a target feature of a system following different strategies.[34,35] Particularly for biomolecules and organic compounds in the CG level, it is possible to find generic classical force fields to describe the interactions between superatoms: these are the MARTINI force fields.[23,24] The form of the force field experienced by each superatom *i*, $F_i = f(r_1,...,r_n)$, within MARTINI includes bonded and non-bonded contributions, being the latter divided into Lennard-Jones (LJ) and, when appropriate, Coulombic interactions. Whilst there are guidelines on how to parametrize each of these contributions,[23,24] the key feature of MARTINI force fields is that they offer a ready-to-go parametrization of the LJ contribution. The latter was developed aiming to properly model the interactions between superatoms formed by different sets of atoms with a same set of parameters. This is achieved by classifying the superatoms by their degree of polarity, hydrogen bonding, amongst other general characteristics, instead of by their precise chemical composition. The influence of how many atoms are lumped into a bead is also taken into account. Notably, the most successful fitting target for the LJ parametrization is, in essence, the partitioning of different macromolecules in a variety of biphasic systems, being the parameters in MARTINI 3.0 refined up to greater lengths to also better reproduce a larger set of thermodynamic properties.

In the present work, MARTINI 2.0 and 3.0 force fields are evaluated concerning their capability of modeling ZIF-8 at the coarse grained level. In doing so, four possible mappings are considered (see figure 1), each with a corresponding indexing for the different bead types. The degree of coarsening decreases going from mapping A to mapping D. It is worth to note that all these mappings are, to one extent or another, out of the scope of the guidelines of MARTINI for designing a mapping. In principle, this would mean deviating from the scenario for which the pre-parametrized force-fields are most optimally applicable to classically model the system. Mappings A and B surpass the recommended number of atoms to lump into a bead, particularly in the context of aromatic rings. The 5-1 mapping of beads type one in mapping C exceeds the recommended degree of coarsening. Its atom splitting is somewhat exotic. Indeed, MARTINI guidelines advocate against splitting chemical groups when designing a mapping, meaning that lumping the nitrogen of the ligands with the $Zn^{2+}$ cations instead of with other atoms of the aromatic ring may not lead to a proper classical modeling of the system when using MARTINI force-fields either. On the other hand, considering a single atom to be a bead, as it is the case in mappings B, C and D, is also not recommended in the MARTINI 2.0 mapping prescriptions. The reason for having considered such mappings regardless is simply to evaluate how much of a detailed description they are able to keep within the scope of MARTINI force fields as well as to allow comparison with models coming from other potential fitting strategies, not restrained by a specific way to map the system.

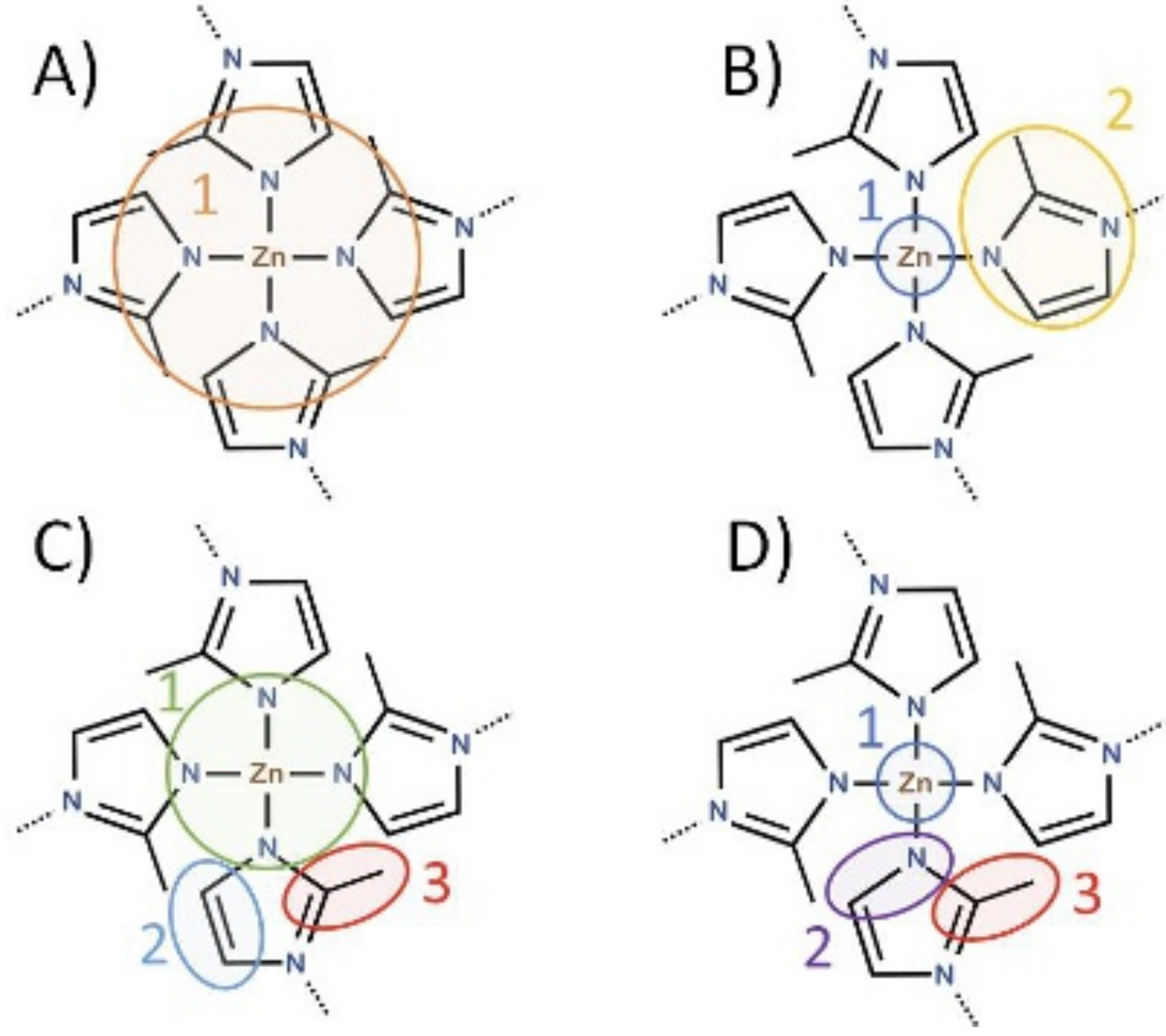

Figure 1. Scheme of the four mappings studied in this work for ZIF-8. Circles indicating the bead assignment for the CG mappings are superimposed to the AA representation of the system: A) mapping A, in which the only bead encapsulates the entire repeating unit; B) mapping B, in which beads type 1 are the $Zn^{2+}$ cations and beads type 2 the entire ligand; C) mapping C, in which beads type 1 comprise the $Zn^{2+}$ cations together with the four nitrogen atoms, beads type 2 the -HCCH- and beads type 3 the $-CCH_3$ groups; and D) mapping D, in which beads type 1 are the $Zn^{2+}$ cations, beads type 2 the -NCH- and beads type 3 the $-CCH_3$.

For each mapping, different classical models resulting from different choices of bead flavors were investigated within both MARTINI versions. Models are referred to as mXMiY where X refers to the mapping (X=A, B, C or D), i refers to the MARTINI version (i=2 or 3) and Y refers to is an arbitrary index for the model in roman numbers (Y=I, II, III, IV or V). The justification for investigating several bead flavors for each mapping lies in the fact that MARTINI force fields were not parameterized for MOFs, making finding the most appropriate choices less evident. Difficulties in setting up bead flavors also arise when many atoms that differ in their degree of polarity are lumped in a single bead, as it is the case in mappings A and B. Different bead flavors, corresponding to different degrees of polarity, were tested in the context of each version of MARTINI for mapping A and mapping B beads. Possibility of charged beads was also considered for mapping B. In all mappings, beads that encapsulate single $Zn^{2+}$ cations were always considered to be either charged or to exhibit a high degree of polarity. For mapping C, the fact that nitrogens are lumped together with $Zn^{2+}$ difficults the determination of the degree of polarity of the overall bead. Thus, nonpolar bead flavors are also investigated for beads type 1 within the two versions of MARTINI. Since the two other bead types within mapping C only contain H and C aromatic atoms, the degrees of polarity considered lie within the apolar range, following the recommendation of bead flavor for aromatic-carbon-based beads. In contrast with mapping C, beads containing nitrogen atoms in mapping D (beads type 2) are allowed a larger degree of polarity as recommended in MARTINI guidelines, so possibilities ranging from non-polar to polar were tested. Beads of type 3, on the other hand, were always considered to be apolar, similarly to mapping C. Naturally, whenever the beads containing the $Zn^{2+}$ cation in mappings B, C and D are charged, the other beads were also assigned to be charged to maintain the overall charge neutrality of the ZIF.

The label "hydrogen bond acceptor" was always considered for beads that contain the nitrogen of $mIm^{-}$. Particularly within MARTINI 3.0, other labels describing additional characteristics of a given bead were also considered. These labels are, occasionally, partial charge, whenever the value of charge of one or more beads fall into the reported range that prescribes the use of the label, and electron polarizability, whenever an attempt to further emphasize electron-concentration of specific beads was made. Initially, superatom sizes were assigned following MARTINI 2 and 3 official publication's guidelines, being the superatom classified as the largest or smallest available size when exceeding or falling short in the amount of atoms it encapsulates, respectively.

Whenever charged bead flavors are considered within a given model, the values of charge for each bead type are determined by summing the charges of the atoms forming it. Values of atomic charge were borrowed from the ZIF-FF model for the all-atom representation.[29] When it comes to the bonded contributions of the MARTINI force fields, only the potentials for bonds and angles were considered. As usual, the connectivity in the coarse grained representation is defined by identifying the superatoms which encapsulate atoms that are chemically bonded to one another. The corresponding equilibrium values of angles and bond lengths for each mapping were obtained from classical molecular dynamics simulations in the all-atom (AA) representation using the ZIF-FF force field. The AA simulations were run in the NVT ensemble at 300K, with the Nose-Hoover thermostat and a damping constant of 100 fs. A time step of 1 fs was used. The duration of the equilibration was 1 ns and the production, 0.5 ns. Microstates were saved during the production and further coarsened by a post processing script according to each mapping individually. Then, angle distribution functions (ADFs) and bond distribution functions (BDFs) were built for all bonds and angles for each of them. The equilibrium values of bond lengths and angles are taken to be the most frequently counted (i.e., most probable) value. In all cases, initial values for the bond and angle force constants were borrowed from the ZIF-FF's potentials. Such values are later optimized to reproduce the peak height and width of reference ADFs and BDFs, as prescribed in the MARTINI guidelines for parametrizing bonded contributions. Notably, for some of the mappings studied, some angles formed by a same set of three superatom types exhibit more than one equilibrium value. Therefore, in order to use the harmonic potential form of MARTINI force-fields without jeopardizing the structural description, it would be necessary to identify these angles and differentiate them as two distinct angle types. Aiming to develop models that can be more readily applicable, potentials for angles falling into this category were initially dismissed.

Before performing the optimization of the force constants of the bonded potentials, the initially set-up models were evaluated concerning their capability to reproduce the structure of the system. This analysis was carried out by comparing RDFs and ADFs with their reference counterparts, taken from the AA simulations. In face of the results, possible modifications in order to improve the quality of the preliminary structural descriptions were investigated for all models. Three possible types of modifications were considered, being each of them implemented only when the structural results of the initially set-up models suggested it to be beneficial: (i) changing the bead size, (ii) implementing bond potentials to regulate the relative positioning of a specific pair of beads that were not bonded judging based on the connectivity criteria (note that this also implies eliminating their non-bonded interactions), (iii) implementing potentials for angles that have more than one equilibrium value. Given the way bead size affects the parametrization, its change is considered in this stage mostly as a tool to regulate the degree of repulsion given by the non-bonded potentials. On the other hand, bonded potentials are implemented as a means of correcting the interspacing between a given set of beads whenever the initial set-up models failed to reproduce it. When potentials for multi-peaked angles are tuned in, the angles are differentiated and treated as different angle types.

After evaluating the possibility of improving the structural description by means of these three possible modifications, the force constants for all bonded potentials within each model undergo optimization. Since the force acting on each given bead is a combination of all existing bonded and non-bonded terms, it is possible to see, within the understanding classical mechanics, that changing one force constant may have an effect not only on the BDF or ADF of the concerned bonded potential, but also, indirectly, on the BDF and ADF of other bonds and angles to one extent or another. This makes it rather complex to craft an automated way to optimize the constants. Thus, the force constants were manually adjusted aiming to reproduce the width and height of the corresponding BDF or ADF whilst avoiding undesirable changes in the other BDFs and ADFs.

All simulations at the CG level carried out for the sake of deriving the structural results were performed in the NVT ensemble at 300K, aiming to match the (T,V) condition considered for deriving the reference structural results in the AA level. The time step was set to 20 fs. Nose-Hoover thermostat was used with a damping constant of 2000 fs, following the recommendation in LAMMPS' manual.[36] Microstates for calculating RDFs, ADFs and BDFs were collected during the production. Equilibration and production lengths varied for different models, being it always sufficient to ensure that the system is equilibrated and that a sufficient amount of microstates were considered to obtain smooth histograms.

Finally, once the values of force constants were settled, the models were assessed concerning their capability of reproducing structure, thermal expansion coefficient ($\alpha_V$), elastic constants and of capturing the guest-induced swing effect, and optimal models were pointed out. Figure 2 summarizes the methodology.

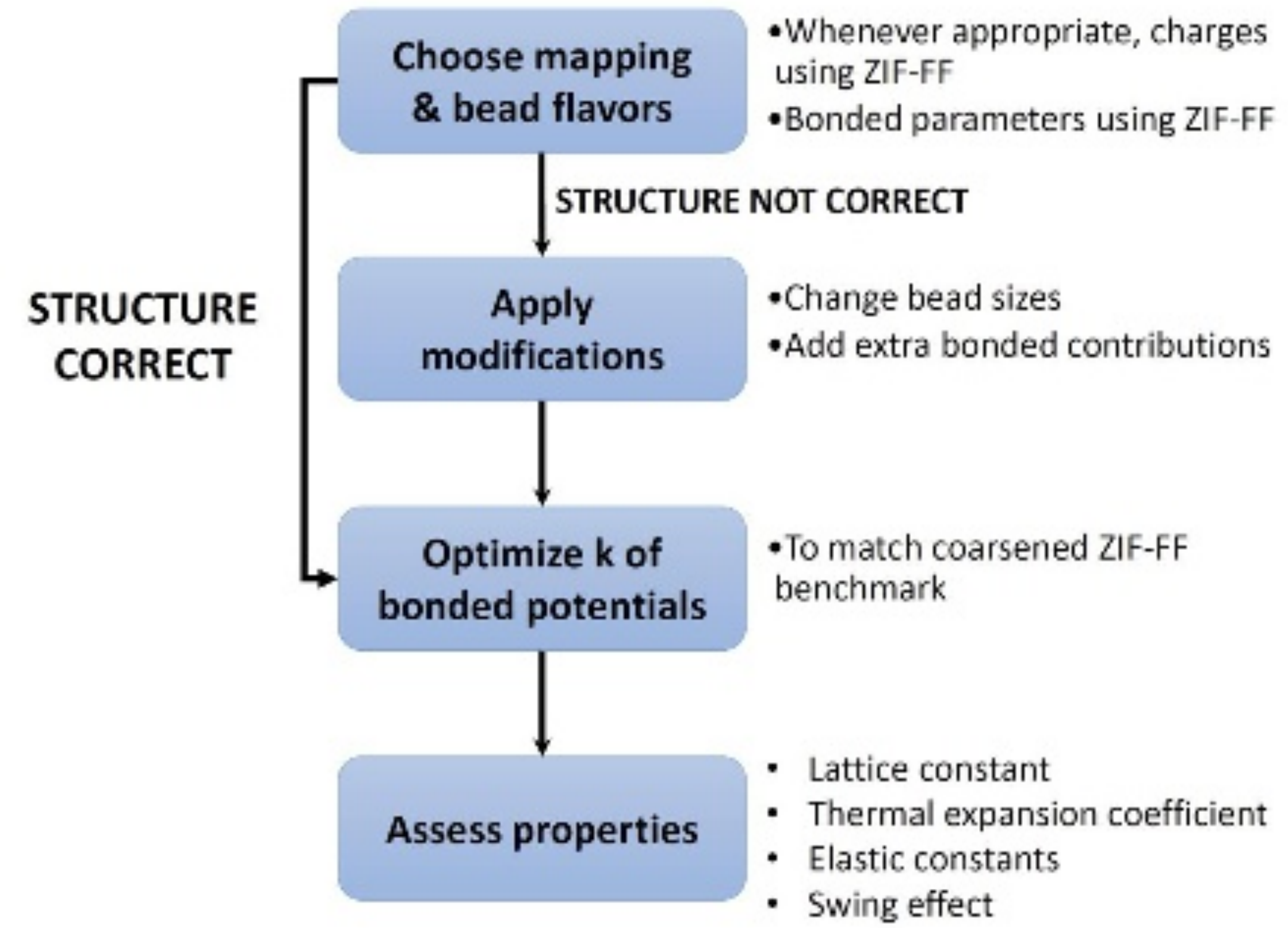

Figure 2. Scheme for the methodology devised for parameterizing the ZIF-8 MARTINI force fields.

## b) Volume expansion coefficient

To get data for estimating the volume expansion coefficient of ZIF-8, MD simulations were made in the NPT ensemble for each model in their respective mapping at 1 atm and two different temperatures: 300K and 272.5K. The derivative $(dV/dT)_p$ was numerically approximated by considering the corresponding volumes at these two temperatures. These simulations were carried out with a time step of 20 fs using the Nose-Hoover thermostat and barostat with damping constants of 2000 and 20000 fs respectively, as recommended in the LAMMPS manual.[36] Equilibration length was adapted for each model as their required equilibration times varied. Simulations at these same (p,T) conditions were also performed in the atomistic scale using the ZIF-FF force-field to derive reference results.[29]

## c) Elastic tensor

Stress and strain are understood to be tensorial entities whose components' correlate during elastic deformation by means of linear relationships, being the coefficients the elastic constants, elements of the so-called elastic tensor.[37,38] In face of the cubic symmetry of ZIF-8's structure, some elastic constants are zero whilst others are equal to one another so that ultimately only $C_{11}$, $C_{12}$ and $C_{44}$ are the relevant ones.[39] Within the present work, the elastic tensor was assessed on two different thermodynamic states: (300K, 0 GPa) and (300K, 0.2

GPa). The pressure here is purely mechanical as the mechanical behavior is investigated for the empty framework.

To determine the elastic tensor at each thermodynamic state, an initial MD simulation in the NPT ensemble was carried out for all models within each mapping to obtain their equilibrium lattice dimensions at that given thermodynamic state. Successively, NVT simulations were carried out in deformed states whose corresponding box's dimensions were altered by either changing the equilibrium length in x direction to generate strains $\varepsilon_{xx}$, or the value of xy to generate strains $\varepsilon_{xy}$. The value of xy corresponds to the x component of the **B** vector, generator of the simulation box within LAMMPS.[40] The two strain types are considered separately and for each of them six different strain values were considered within the interval [-0.6%, 0.6%]. At each deformed state, the macroscopic values of different stress components, $\sigma_{ij}$, were determined from averaging their instantaneous values calculated and printed by LAMMPS. The values of $C_{11}$, $C_{12}$ and $C_{44}$ were determined by finding optimal linear fits for $\sigma_{xx}=f(\varepsilon_{xx})$, $\sigma_{yy}=f(\varepsilon_{xx})$, $\sigma_{zz}=f(\varepsilon_{xx})$, $\sigma_{xy}=f(\varepsilon_{xy})$, respectively. Particularly, the value of $C_{12}$ is taken to be the average of the coefficients obtained in both $\sigma_{yy}=f(\varepsilon_{xx})$ and $\sigma_{zz}=f(\varepsilon_{xx})$ linear fits. The set-up methodology is in essence the explicit volume deformation method reported in the literature.[41]

Finally, the capability of the models of reproducing the amorphization of ZIF-8 was evaluated by running MD simulations at (300K, 0.4 GPa), the thermodynamic conditions in which the phase transition should take place.[42] The simulations were carried out in the NPT ensemble for a total of 200 ns (note that the timestep used is of 20fs). RDFs between all bead pairs were built every 50ns to evaluate possible structural changes.

Barostat and thermostat specifics and time step were the same as in section b.

## d) Presence of guests: swing effect and pairwise guest-MOF interaction

The swing effect consists of a subtle phase transformation that occurs upon guest sorption together with the corresponding mechanical stimuli at different temperatures.[43-46] The new phase, referred to as high-pressure (HP) phase, differs from the ambient pressure (AP) one by a certain degree of rotation of the $mIm^-$ rings as well as a slightly higher unit cell length (16.98Å and 17.07Å respectively, as reported in the context of MeOH sorption).[43] The orientation of the linkers can be described by the value of the swing angle, denoted as $\varphi$, which corresponds to the dihedral angle Zn-Zn-Zn-$CCH_3$ shown in figure 5(a), which exhibits different values in the AP and HP structures (≈7° vs ≈35°).[47] In the AP structure the linker is more in-plane with the 6MR window thus reducing its effective size compared to the HP structure. It is important to note that each linker is part of one 4MR window and two 6MR windows and therefore, naturally, given that the two 6MR windows are not in the same plane, the swing angle cannot be of ≈7° (or ≈35° if in the HP structure) for every single dihedral depicted in figure 5(a).

For the model to reproduce the swing effect, it would be necessary that it predicts the HP structure as being the stable one in the regions of adsorption isotherms in which this phase has been identified as an occuring phase. The interactions within the system should allow for a new stable state, marked by the proper degree of rotation of the linkers and an increase in lattice parameter at the rightful thermodynamic states. The methodology used to assess this in the present work consists of carrying out MD simulations in the NPT ensemble at a given temperature, guest loading and its corresponding pressure in which the HP phase is the stable one. The resulting value of the swing angle of the equilibrated structure is then evaluated.

The guests considered are methanol (MeOH) and $N_2$ with loaded states of 41 and 51 molecules per unit cell respectively, with corresponding pressure values taken from experimental adsorption data for MeOH at 300K and for $N_2$ at 77K.[43,44] The gasses were considered in their AA representation as rigid bodies with interactions between molecules classically modeled *via* the united atom TraPPE force field.[48,49] The gas-MOF parameters for the LJ pairwise interactions were established by the Lorentz-Berthelot mixing rules for each model. For these MD simulations, the time step was of 1 fs, smaller than the values for which MARTINI models were standardly calibrated, aiming to guarantee simulation stability at such high loadings. The cutoff adopted for all the LJ and Coulombic interactions are the ones prescribed within the MARTINI force-field for each model, namely 12 Å for models coming from version 2.0 and 11 Å for those coming from version 3.0. Pairwise interactions between all pairs were set up to follow the shift strategy prescribed in the MARTINI guidelines. The framework is initialized in its AP structure containing the respective amount of molecules per unit cell for each gas separately. Damping constants of 100 fs and 1000 fs were used for the Nose-Hoover barostat and thermostat respectively. Equilibration time varied according to the needs of the different models. Microstates were saved during the last 0.5 ns for calculating the swing angle distribution function through a post-processing script. The possibility of capturing the swing effect was investigated only for mappings C and D, which are the less coarsened ones. The results are compared with the ones obtained for the empty structure in the context of each respective model. Further investigation was pursued guided by the results' interpretation.

The overall interaction energies between the superatoms of the MOF and the nitrogen molecules were also calculated for the mappings C and D models, aiming to analyze the influence of bead flavors within the two MARTINI versions in the guest-host interactions. This calculation consists of the summation of the interaction energies between the atoms of each and all $N_2$ molecules and all the MOF superatoms of a specific type that lie within a distance equal or smaller than a cutoff distance from them. The summation is done for different microstates and averaged afterwards. The calculation was made for all superatom types within mappings C and D. The interaction energy encapsulates the Lennard Jones interactions and, for models in which the MOF is charged, Coulombic interactions. The cutoff distance used as criterion for deciding whether or not to include an interaction energy in the analysis is the same as the one prescribed within MARTINI for pairwise interactions, namely 12 Å for M2 models and 11 Å for M3 models. Notably, the fact that the cutoff used as criteria is different for models coming from different MARTINI versions allows for more pairwise interactions to be counted in M2 models, thus making it less straightforward to compare the impact of bead flavors on the interaction energy values between MARTINI versions. Aiming to counter this issue, the average interaction energies were normalized by the average amount of ($N_2$ molecules)-(MOF superatom) pairs that were involved in the calculation to ultimately yield the corresponding pairwise interaction energy value. Note that these energies should not be linked to any experimental quantity nor to binding energies. Yet, these values of energy can still be regarded as an interesting metric on how the bead flavor within MARTINI models can influence the guest-host interactions to make them more or less thermodynamically favorable.

# RESULTS AND DISCUSSION

## a) Structural results and parameterization

All the initially set up MARTINI models for mapping A properly reproduce equilibrium neighbor distances as well as equilibrium angle values. However, implementing potentials for angles was required to reproduce peak width in the ADFs profiles. All mBM2 and mBM3 models required implementing potentials for angles 212 in order to capture the existence of all peaks in RDFs and ADFs as well as their alignment. A further reduction of the bead sizes was also necessary in order to have more rightful peak alignments in RDFs and, most importantly, on ADFs, as increasing the force constants of the bonded potentials merely led to a reduction of the amplitude of motion.

For what concerns mCM2 and mCM3 models, the bonded contributions initially considered were not sufficient to properly regulate the relative spacing between superatoms 2-2, 2-3 and 3-3. As the most drastic failing point was the severe underestimation of first 2-3 neighbors, a bonded potential regulating the distance between superatoms 2-3 within a ring was first created. However, solely implementing it was not sufficient to solve, by indirect effects, the poor description of ring-ring relative positioning for neither of the two MARTINI versions. All the models continued to underestimate, to one extent or another, the first neighbors 2-2 distance and to overestimate the first neighbors 3-3 distance. This led to further implementation of angle potentials for the 212, 313 and 312 angles, which had been initially dismissed. Figure 3(a), (b), (c) portrays the evolution of the structural description in terms of the 2-2, 2-3 and 3-3 RDFs for the mCM2III model, arbitrarily chosen, as each of the implementations mentioned are made.

Lastly, for mapping D, the initially set-up classical mDM2 models did not yield numerically stable MD simulations - bonded atoms found themselves way too far apart. Aiming to gather more information about the issue, $K_{bond}$ constants were diminished and potential for angles were dismissed to get a glimpse on the resulting spatial organization of the superatoms. Upon judging the results, the problem seems to be an excessive repulsion coming from the non-bonded potentials that causes unrealistically large forces to arise as the bonded contributions try to bring closest neighbors closer together to set up their rightful relative positioning. Figure 3(d) shows the 1-1, 1-2 and 2-2 RDFs of the structural results obtained for model mDM2IV. The modifications sequentially made to counter the problem included diminishing bead size, adding a potential for angles 212 that had been initially dismissed and even excluding the LJ interactions between the closest 1-3 neighbors together with implementing a potential for bonds to try to properly regulate their spacing. Particularly, implementing the last two modifications simultaneously did not yield MD simulations that are numerically stable for any set of $K_{bond}$ and $K_{angle}$ values for the potentials for bond 1-3 and angles 212, causing one modification to have to be chosen over the other.

Ultimately, regardless of all efforts, all mDM2 models continued to overestimate most of the distances between neighbors, suggesting that even after the diminishment of bead size, the non-bonded potentials are still excessively repulsive and predominate over the bonded contributions. In contrast, mDM3 models showed better structural results. No diminishment of bead sizes was necessary, which can be traced back to the fact that the latter version comes with LJ parameters that are better parametrized for mappings that are not so highly coarsened.[24] Potentials for 212 angles and for regulating the 1-3 first neighbors distance were both required though, to achieve a better structural description, being it possible to implement both modifications simultaneously in this case.

Once the modifications were implemented for all models within all mappings, the force constants of the bonded contributions were optimized. The final set-up of bead flavors corresponding to each model is shown in table 1. Figures showing the structural description, given in terms of ADF, BDF and RDF, of the final models can be found in the Supplemental Material (SM) (figures SM1-SM8). The detailed values of their non-bonded and bonded parameters can also be found in the SM (tables SM1-SM3). It is interesting to note that the different models coming from the same MARTINI version result in structural descriptions that are quite alike to one another, revealing that the choice of bead flavor is not so decisive for the depiction of the structural arrangement of the atoms for the empty MOF. Additionally, it is possible to note that ultimately the values of the optimized force constants for a same bonded potential do not vary so much between models coming from the same version of MARTINI for a given mapping.

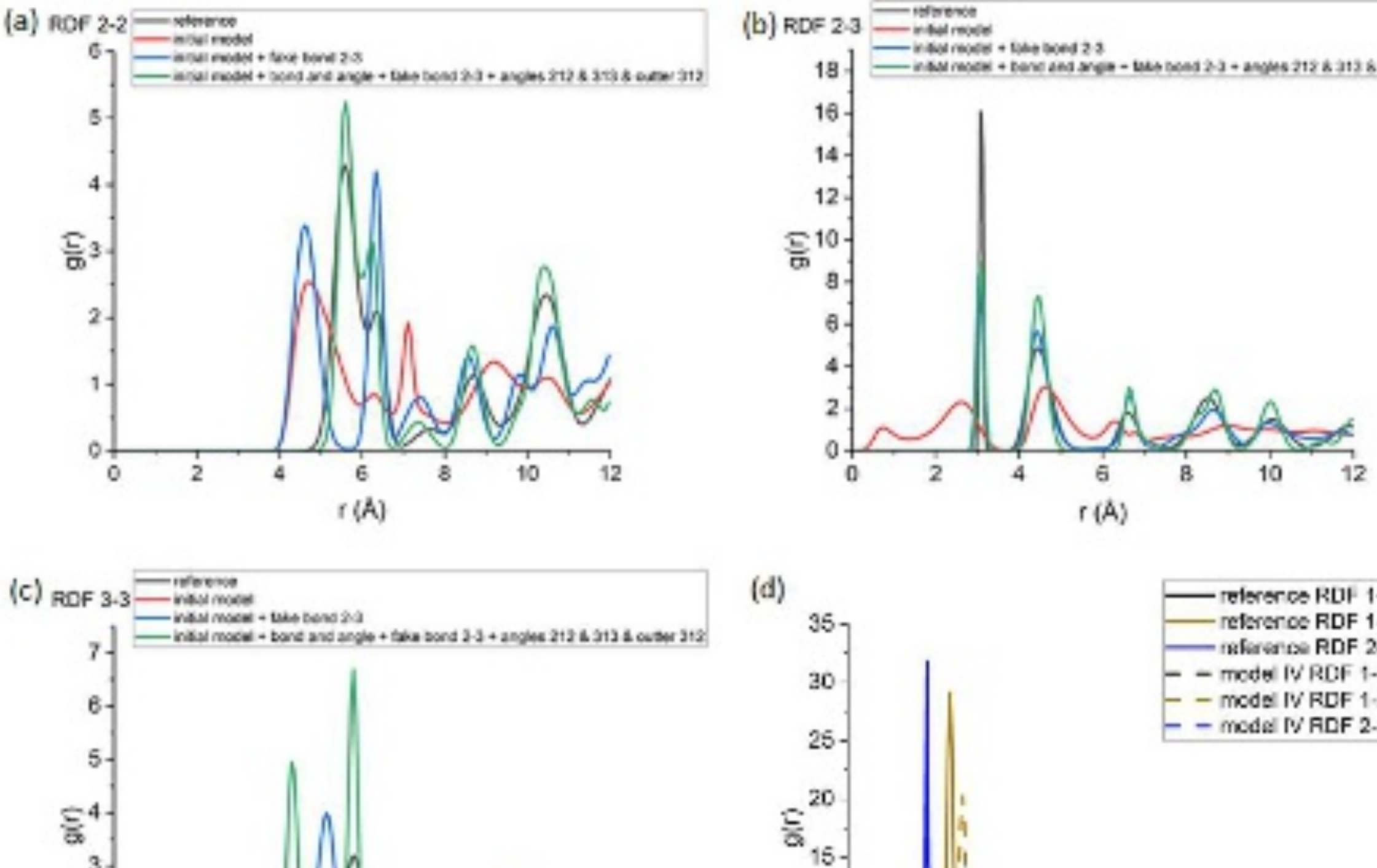

Figure 3. Evolution of the relative spacing of superatoms (a) 2-2, (b) 2-3 and (c) 3-3 as modifications are implemented in model mCM2III; (d) 1-1, 1-2 and 2-2 RDFs of the coarsened atomistic reference and of the mDM2IV model in which potential for angles are dismissed and $K_{bond}$ values diminished.

Particularly for mDM2 models, which had revealed themselves to be susceptible to numerical stability issues in MD simulations, 50 ns long MD runs in the NPT ensemble at ambient conditions were carried out with the optimized parameters. Although there is no way to certify that the models are indeed numerically stable, such a long term run reveals that, at least within that large amount of time steps, the MD simulations are stable at ambient (T,P) condition. These simulations were run with a 20 fs time step, Nose-Hoover thermostat and barostat with values for the damping constants as recommended by LAMMPS.[36] Among the four models, model mDM2IV was the only one that continued exhibiting numerical instability issues during the dynamics, causing it to be excluded from further analysis. The instability may be linked to the continuous fight between the bonded and the overly-repulsive non-bonded contributions of the force field, which was not possible to solve by the modifications implemented in the earlier stage. Indeed, the epsilon of LJ potentials for model mDM2IV are overall higher compared to the others, supporting the previous reasoning.

| | I | II | III | IV | V |
|---|---|---|---|---|---|
| **Mapping A MARTINI 2.0** (mAM2) | Head type 1: C3 | Head type 1: $N_a$ | Head type 1: P3 | | |
| **Mapping A MARTINI 3.0** (mAM3) | Head type 1: RC3 | Head type 1: $HN2_a$ | Head type 1: $HP3_a$ | | |
| **Mapping B MARTINI 2.0** (mBM2) | Head type 1: Q0<br>Bead type 2: $Q_a$ | Head type 1: P5<br>Bead type 2: P5 | Head type 1: P5<br>Bead type 2: $N_d$ | Head type 1: P5<br>Bead type 2: C4 | |
| **Mapping B MARTINI 3.0** (mBM3) | Head type 1: $SP5_q$<br>Bead type 2: $RN2_{aq}$ | Head type 1: SP5<br>Bead type 2: $RC3_e$ | Head type 1: $SP5_q$<br>Bead type 2: $RP2_{aq}$ | Head type 1: SD<br>Bead type 2: $RQ1_n$ | |
| **Mapping C MARTINI 2.0** (mCM2) | Head type 1: $Q_a$<br>Bead type 2: SQ0<br>Bead type 3: SQ0 | Head type 1: P5<br>Bead type 2: SC1<br>Bead type 3: SC1 | Head type 1: $N_a$<br>Bead type 2: SC1<br>Bead type 3: SC1 | Head type 1: P5<br>Bead type 2: SC5<br>Bead type 3: SC5 | Head type 1: $N_a$<br>Bead type 2: SC5<br>Bead type 3: SC5 |
| **Mapping C MARTINI 3.0** (mCM3) | Head type 1: RQ3<br>Bead type 2: TQ1<br>Bead type 3: TQ1 | Head type 1: RD<br>Bead type 2: TQ1<br>Bead type 3: TQ1 | Head type 1: HP5<br>Bead type 2: TC6<br>Bead type 3: TC6 | Head type 1: $HP5_a$<br>Bead type 2: TC6<br>Bead type 3: $TC6_v$ | Head type 1: $RP5_{qa}$<br>Bead type 2: TC6<br>Bead type 3: $TC6_q$ |
| **Mapping D MARTINI 2.0** (mDM2) | Bead type 1. SQ0<br>Head type 2: $SQ_a$<br>Head type 3: SQ0 | Bead type 1. SP5<br>Head type 2: SP5<br>Head type 3: SC1 | Bead type 1. SP5<br>Head type 2: $SN_a$<br>Head type 3: SC1 | Bead type 1. SP5<br>Head type 2: SP5<br>Head type 3: SC5 | Bead type 1. SP5<br>Head type 2: $SN_a$<br>Head type 3: SC5 |
| **Mapping D MARTINI 3.0** (mDM3) | Bead type 1: SD<br>Bead type 2. $TQ1_n$<br>Bead type 3. TQ1 | Bead type 1: SP5<br>Bead type 2. $TN3_a$<br>Bead type 3. SC6 | Bead type 1: SP5<br>Bead type 2. $SN3_a$<br>Bead type 3. $SC6_v$ | Bead type 1: $SP5_q$<br>Bead type 2. $TN3_{aq}$<br>Bead type 3. $TC6_q$ | |

Table 1. Final bead flavors of each underlying model considered within both versions of MARTINI for each mapping.

## b) Physical and Mechanical properties

As a continuation of our study, the thermal expansion coefficient, lattice parameter at 300K and elastic tensor were estimated for all the final models within the respective mappings. The results can be found in figure 4,  together with their standard deviations. The values of lattice constant were obtained from the NPT simulations at (300K, 0 GPa), previously described in the elastic tensor subsection of the methodology.

During the assessment of elastic constants at (300K, 0.2 GPa) some models revealed themselves to be numerically unstable at the given thermodynamic state and, as a consequence, results for these models are not reported in figure 4. All mDM2 and mDM3 models and the non-charged MARTINI 3.0 models for mapping mCM3 (mCM3III, mCM3IV, mCM3V) show instabilities during the equilibration of the system at (300K, 0.2 GPa), always within the first 60 ns of simulation time (note that the timestep used is 20 fs). Exceptionally, models mDM3II and mDM3III were successfully equilibrated at such thermodynamic state, but during the equilibration at deformed states, the simulation collapses. In all cases, this happens due to bonded superatoms being launched too far apart from each other. The reason for the instability is believed to be the inability of the coarse grained models to reproduce the dynamics in a state where the superatoms are forced into such smaller interspacing with correspondingly higher values of forces experienced, as implied by the compressive pressure condition. The shortening of the distances, combined with the rather large timestep considered, could cause unrealistically large repulsive forces which, ultimately, when assumed to last for the entire duration of the timestep, lead to setting bonded pairs farther apart than fluctuations should allow.

More disconcertingly, mCM3III, mCM3IV, mCM3V and mDM3II exhibited numerical instabilities during the equilibration at isolated deformed states considered for collecting stress vs strain data to estimate the elastic constants at (300K, 0GPa). The cause for the simulation collapse is also bonded superatoms being found too far apart from one another. While properties for unstable models can still be estimated upon considering the successful phase space trajectories, this raises a red flag concerning their reliability in modeling interactions at all. In response to the instability of the models under question in modeling the dynamics of coarsened ZIF-8 at ambient conditions, mCM3III, mCM3IV, mCM3V and mDM3II are discarded as valid CG models for ZIF-8. Their physical and mechanical properties in such conditions are hence not shown in figure 4.

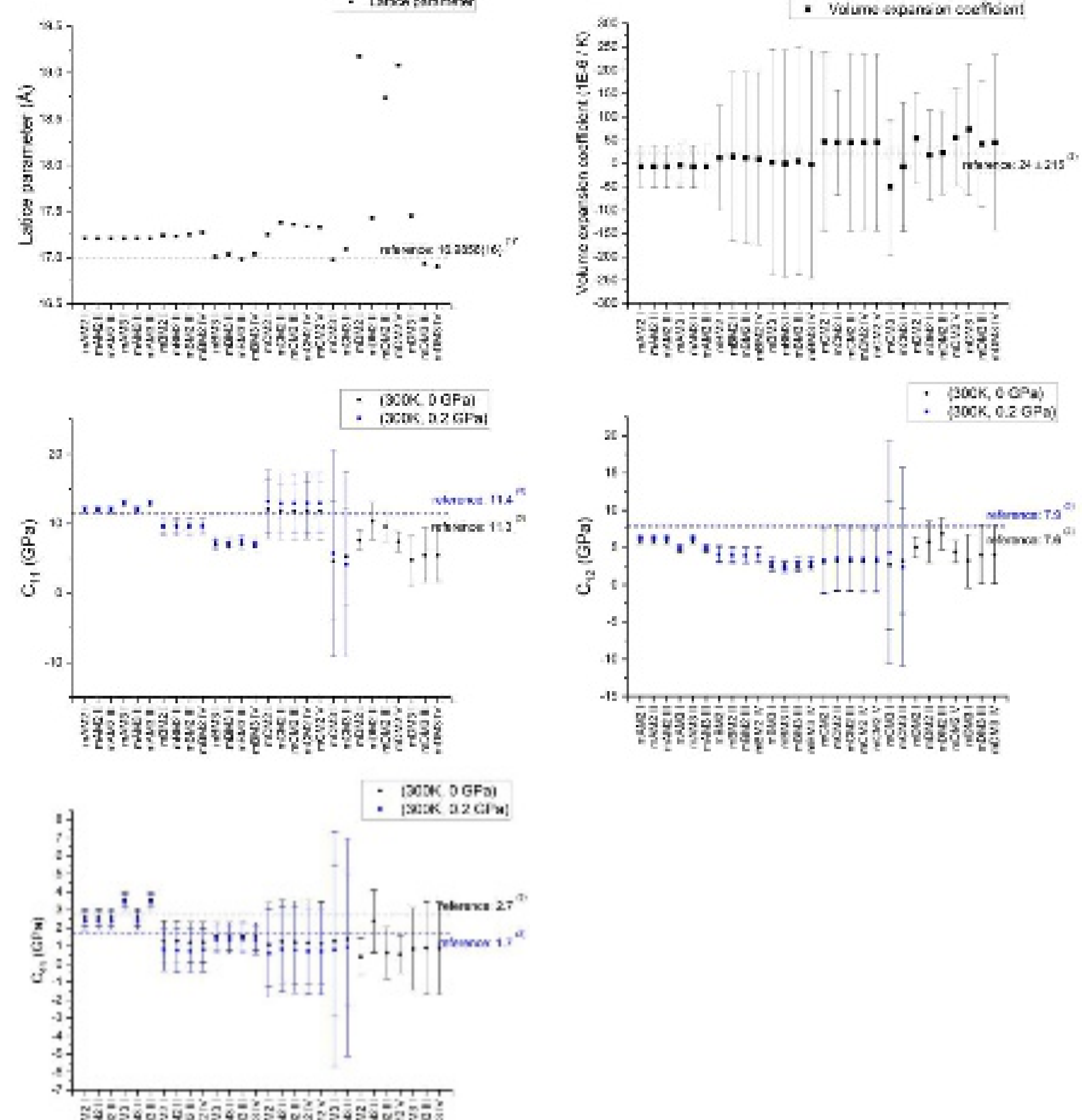

Figure 4. Results on (a) lattice parameter, (b) volume expansion coefficient, (c) $C_{11}$, (d) $C_{12}$ and (e) $C_{44}$ elastic constants of ZIF-8 calculated for each MARTINI model under assessment for the different mappings. (1) Ref. 43 (experimental data), * at ambient pressure (2) ZIF-FF potential, (3) Ref. 42.

On what concerns lattice parameter, it is possible to conclude that the experimental reported value (16.9856 Å) is overall well reproduced by most of the models. MARTINI 2.0 models show however a tendency to overestimate it compared to their MARTINI 3.0 counterparts. The most drastic overestimations (of about 12%) are the values given by models mDM2I, mDM2III and mDM2V. This is in alignment with the general high repulsion tendency pointed out in the previous section for these models.

The thermal expansion coefficient measures the change in volume of the system when changing its temperature, being the temperature tied algebraically roughly exclusively to the kinetic energy of its structural units. It is possible to see in figure 4(b) that some models tend to predict negative values for it. To grasp these results, it is important to understand how volume expansion is addressed within classical mechanics, which is afterall the formalism that was used to derive the dynamics of the system in the present context. In the coarsened systems under examination, all superatoms are under the effect of force-fields of asymmetric well-shape since, besides the symmetric bonded contributions coming from the harmonic potentials, they all interact also via asymmetric 12-6 LJ non-bonded interactions with their neighbors. After some reasoning within the mathematical formalism of classical mechanics, it is possible to conclude that a same net change, caused by the force-field, on the velocity ($|\Delta v|$) of an oscillating particle will always take more time to occur on the softer side of the potential compared to the sharper one. It follows as a consequence that the particle is fated to spend more time of its overall dynamics in the less steep side of the potential. Notably, this becomes more and more accentuated as the amplitude of the motion increases.

For an equilibrated multi-particle system at a given thermodynamic state, the resulting average relative distance between two given neighbors can be identified in the RDF results. Table SM4 summarizes the average spacing of the closest pairs that interact via non-bonded interactions at (300 K, 1 atm) as well as, for each respective pair, the value of $R_{min}$ of the LJ potential dictating their interactions. The latter is equal to $\sigma\sqrt[6]{2}$, value for which the force coming from the LJ interaction is zero.

In the context of the motion within a potential well, an increase in kinetic energy of the oscillating particle implies an increase in the amplitude of its motion. This, in the context of an asymmetric potential and of the discussion previously made, ultimately leads either to a diminishment or an increase of the relative distance between neighbors depending on which side is the soft one (left- or right- hand side with respect to the equilibrium position).

For all MARTINI models for mapping A, the spacing between closest pairs of neighbors of all types that interact via LJ potentials is larger than $R_{min}$ for the given LJ field. This means that for every single superatom, the resulting force field, obtained upon summing the non-bonded and bonded contributions, will be an asymmetric well with a smoother left-hand side. This supports the notion that the continuous increase in kinetic energy will continually shorten the relative distances between neighbors, thus supporting the shrinking of the system as the average kinetic energy increases, namely a negative thermal expansion coefficient. The same holds for the mCM3 models. Indeed, data collected from successful trajectories reaching up to 16 ns reveal values of $\boldsymbol{\alpha}_v$ of -35.23, -3.86 and -7.42 /MK for models mCM3III, mCM3IV and mCM3V respectively, which are in agreement with the reasoning made.

Notably, thermal expansion results also support the formulated theory about why the non charged mCM3 models in particular have correspondingly numerically unstable MD simulations in the long term at both 0.0 GPa and 0.2 GPa. For these models, forces coming from the non-bonded potential lie unanimously in the attractive region, fighting to bring the superatoms closer together during the motion. As a response, high repulsive bonded forces, which could eventually launch the atoms too far from one another, may arise. Notably, because the LJ force-field between some superatom pairs for mDM3II lie in the attractive range, it is likely that this is also the underlying reason behind its instability at ambient condition. Interestingly, the scenario depicted for the non-charged mCM3 models and mDM3II are opposite to that of the mDM2 models. These considerations put together reinforce the importance of having balanced non-bonded and bonded potential degrees of repulsion/attraction for obtaining a numerically stable model within the scope of a given equilibrium spatial arrangement of the system. This shall be particularly important in the context of the high timesteps generally used in CG simulations.

For the remaining models for other mappings, it is possible to see that the different pairwise interactions have $R_{min}$ values which are sometimes larger and sometimes smaller than the relative distance between the corresponding pair of superatoms in the resulting structure of the model. As a consequence, it is not possible to identify in a straightforward manner the tendency of the behavior of the system as its kinetic energy is increased. In these cases, the calculated volume expansion coefficients themselves can be regarded instead as an indicator of what contribution is the most predominant. Among the positive values, there is no clear tendency for models coming from a given MARTINI version for a given mapping to give better results. Models mDM2II and mDM2III shield values closest to the reference.

In terms of mechanical properties, all models yield larger $C_{11}$ than $C_{12}$ values, albeit to a lower extent in the case of mDM3 models. No single model exhibits the most accurate values of $C_{11}$, $C_{12}$ and $C_{44}$ simultaneously. Fixing the analysis at (300K, 0 GPa), the value of $C_{11}$ is better reproduced by mBM2 and mCM2 models whilst $C_{12}$ is better reproduced by, overall, mAM2 and mDM2 models. The value of $C_{44}$ at that same thermodynamic state is better reproduced by mDM3 models. Finally, upon looking at the overall results of elastic constants, it can be concluded that within a same mapping, models coming from the same version of MARTINI tend to have similar behavior under elastic deformation. In other words, bead flavor does not seem to play an important role in the elastic tensor.

Concerning response to shear stress at different pressures, all the models having numerically stable simulations at 0.2 GPa predict a decrease in $C_{44}$ compared with the 0 GPa value at 300K, feature linked in the literature to the amorphization of ZIF-8.[42] However, when evaluating the capability of depicting the amorphization of the system, none of the models showed evidence of being able to do so at (300K, 0.4 GPa) within the simulation time considered. mA, mB, mCM2 and mCM3II models yielded RDFs that essentially maintain the peak profiles characteristic of ZIF-8's crystal structure. The peaks were simply slightly shifted to the left compared to the reference RDFs collected at ambient (T,V) conditions for ZIF-8, as expected from simulating the system at a higher pressure. At (300K, 0.4 GPa), mCM3I exhibited the above-mentioned numerical instability issue of bonded pairs being found too far apart, which can be explained with the same reasoning as for those observed at (300K, 0.2 GPa). None of the models that revealed numerical instability issues at (300K, 0.2 GPa) were tested concerning their ability of depicting the amorphization since the numerical instability issue should perpetuate itself at even higher pressures.

Notably the values for the standard deviations of elastic constants and volume expansion coefficients are quite high. When doing the statistics, the high values of standard deviation in the elastic constants originate from the magnitude of the standard deviation associated to the values of $\sigma_{ij}$ at the deformed states investigated. For the thermal expansion coefficients, the high standard deviation comes from the high standard deviation of the difference in volume, obtained during the propagation of errors, at the two temperatures considered for evaluating volume expansion. Notably, this is observed despite the system being equilibrated. Implementing other barostatting strategies may reduce fluctuations in instantaneous values of properties.

It may be interesting to remember, when thinking about the underlying physical cause behind volume expansion and elastic deformation, that modeling the system at the coarse grained level makes it impossible to address the two phenomena in its real origins since the impact of the atoms lumped into a bead is lost. At the coarse grained level, it is the interactions between the superatoms upon changing the temperature or upon mechanically stretching and compressing the material that are, instead, responsible for the values of these properties. In that sense, and given that the MARTINI force-fields were not parameterized for MOFs, the performance of the MARTINI models in estimating thermal expansion coefficients and elastic constants is quite reasonable.

## c) Swing effect and pairwise guest-MOF interaction

mDM3II, mCM3III, mCM3IV and mCM3V, which had been previously discarded from the analysis of physical and mechanical properties due to numerical instability at (300K, 1 atm), were kept for assessing the swing effect on as the presence of guests may balance the attraction between the superatoms of ZIF-8 coming from the LJ interactions. Notably, as previously mentioned in the methodology section, two equilibrium values must exist for the swing angle. The results here reported concern only those formed by linkers and 6MR windows which are close to being in-plane within the initial empty structure.

mCM2 models do not show any significant shift of $\varphi$ upon guest adsorption, which remains on average constant at 14.5° for all models regardless of the structure being empty or loaded with $N_2$ or MeOH (see Figure SM9). mCM3 models have a swing angle value of 0.5° in the empty structure at ambient pressure. Notably, the difference in swing angle values for the empty framework for models coming from the two MARTINI versions could be due to a lower LJ repulsion in the MARTINI 3.0 version, allowing the linkers to be slightly closer, as implicated in the alignment with the 6MR window plane. At a loading of 51 $N_2$ molecules per unit cell, the peak initially at 0.5° shifts to around 8° for all models. Particularly, the non charged mCM3 models (i.e., models mCM3III, mCM3IV and mCM3V) start exhibiting an additional peak centered around the corresponding swing angle value of the HP structure ($\approx$35°). The fact that this peak appears solely for the non-charged models suggests that electrostatic interactions are actually countering the swing effect by overly structuring the system. In the case of MeOH loading for these same non-charged models, it is possible to spot the formation of a shoulder in the swing angle's distributions at larger values. Figure 5(b) shows the distributions for the swing angle obtained with the model mCM3V upon loading and swing angle distributions for other mCM3 models can be found in the SM (Figure SM10). mDM2 models have histograms of swing angle that are too noisy to lead to statistically meaningful conclusions regardless of the MOF being empty or loaded (see Figure SM11). Finally, mDM3 models have the same value of swing angle of 0.5° when loaded with $N_2$, MeOH as well as when empty (see Figure SM12).

Ultimately, it is possible to conclude that all models failed to reach the prescribed stable structure for this level of loading of both $N_2$ and MeOH as the peak at 35° either does not exist or has a significantly small amplitude. Notably, the atomistic force field ZIF-FF cannot reproduce it either, as shown by an attempt to investigate its capability of capturing it for $N_2$ and MeOH as guests following the same methodology as for the CG models. This shows that the challenge lies beyond a matter of the scale used to represent the system. The swing angle phase transition has only been captured via *ab initio* molecular dynamics approaches up to the moment of this writing.[47]

Aiming to further investigate the nature of the results obtained with the non-charged mCM3 models, simulations in the NVE ensemble for both gasses were performed to evaluate whether or not the results can be influenced by barostatting. Pressure fluctuations have after all been reported to be a possible cause for not being able to reproduce dynamical properties that are correlated to cell fluctuations.[50] The volume was set as the equilibrium value obtained in the NPT simulations for each model respectively aiming to evaluate precisely the same thermodynamic state. The results reveal that indeed the same thermodynamic state is successfully achieved in the two ensembles as average values of thermodynamic properties are the same. No differences in the profile of the swing angle's distribution were observed between the two ensembles, discarding thus the possibility of the results being influenced by the barostat used.

In a further investigation, the MD simulations for mCM3 non-charged models, in which a partial reorientation of the swing effect was observed, were run with the 51 $N_2$ molecules per unit cell loading using as initial microstate the HP structure. The goal was to see whether or not the convergence to the final equilibrium state reached would be influenced by the initial microstate used to run the MD simulations. Interestingly, the histograms of swing angle values revealed that the peak at ≈35° completely disappeared when using mCM3III and mCM3IV models, and diminished greatly in the case of model mCM3V. This can be seen in figure 5(c). This reveals that the initial microstates used to run the investigation can have an impact on the final state that is reached. In other words, the system can be trapped in different thermodynamic states, marked by a different set-up of orientations of the linkers. Counterintuitively, it is interesting to note that it doesn't really seem that initializing the system in the HP structure makes it at all more propense to have a final state containing more linkers which are rotated. The influence of the initial microstate used is another red flag on the inability of the models to reproduce the swing effect during classical dynamics within the total simulation time considered.

Guided by further desire to understand the results, grand-canonical Monte Carlo (GCMC) simulations were run to see what would be the equilibrium amount of $N_2$ adsorbed in the rigid framework, modeled in the CG level, specifically at (77K, 1 bar) and (77K, 50 bar). The goal was to see if the computational isotherms obtained using the mCM3III, mCM3IV and mCM3V models were shifted with respect to the experimental one. If the hypothesis was confirmed, it could be that the HP structure is indeed predicted by the models as the stable one but higher pressure and/or loading conditions would need to be investigated. The GCMC was made using the HP structure with a unit cell length of 17.07 Å as no precise information concerning the value of unit cell length of the HP structure at that stage of $N_2$ loading was found. The simulation revealed an equilibrium uptake at 1 bar smaller than 51 molecules per unit cell for all three models (41.7, 42.4 and 41.2 $N_2$ molecules per unit cell for mCM3III, mCM3IV and mCM3V, respectively). Increasing the pressure up to 50 bar did not significantly increase the equilibrium uptake, proving that the saturation capacity of the HP structure in the eyes of the MARTINI models used had already been reached.

Notably, the equilibrium uptake at (77K, 1 bar) is closer to the one obtained in previous works using the AP structure (simulated in full atomistic scale).[44] Further GCMC simulations were carried out with the rigid AP structure at that same thermodynamic state using models mCM3III, mCM3IV and mCM3V and revealed a similar value of equilibrium adsorbed molecules as for the rigid CG HP structure (44.7, 44.6, 44.8 $N_2$ molecules per unit cell for models mCM3III, mCM3IV and mCM3V, respectively). The results suggest that the $N_2$ rearrangement inside the pores and the existence of new sites, proposed in the literature to occur in the HP as an explanation of its larger guest molecules uptake, are simply not captured in the present CG scale models.[44] Additionally, the values of internal energy for both loaded systems (i.e., AP framework + $N_2$ and HP framework + $N_2$) are close to one another, with the AP structure being slightly more energetically favorable (values of energy ranging from 5% to 8% smaller). Notably, internal energy contributes in the general expression that dictates stability of a phase relative to another within classical thermodynamics. Thus, comparing the values of internal energy suggests that the AP structure is more stable than the HP one under this loading condition within these classical models, in clear contrast with what would be needed to capture the swing effect.

Finally, aiming to give a perspective on what could be done in order to capture the swing effect in the coarse grained level, changing the ε values for all cross LJ interactions (guest-host) was tested, inspired in previous work by D. Fairen-Jimenez *et al.*[51] This was made by increasing and decreasing by a given percentage the values of ε, originally obtained by the Lorentz-Berthelot mixing rules. Only the mCM3 non-charged models with $N_2$ as a guest were considered for this investigation. The results show that the peak for the swing angle at ≈35° increases its amplitude with increasing ε values. Figure 5(d) shows the results for model mCM3V. However, by drastically increasing the value of ε, the equilibrium value of the swing angle starts increasing and a non-realistic third peak sitting at around 50° also appears. This "dead end" of the increase in ε of cross LJ interactions in capturing the swing effect suggests that the increase in epsilon has to be made carefully and has to be coupled with other modifications. Notably, to evaluate once again the possible influence of the initial microstate used, simulations were made following the same methodology but having the HP structure as the initial microstate. These simulations revealed a different final structure. Similarly to what was observed for the unmodified potentials, the intensity of the peaks sitting at ≈35° diminished when starting the simulation with the HP structure. However, despite this diminishment, the peak corresponding to the HP structure is still larger compared to what was observed for the unmodified potential, suggesting that adjusting ε of cross interactions could indeed be the key to depicting the effect independently of the initial microstate used.

Lastly, the same strategy of increasing ε of guest-host cross interactions was tested for mCM2 models. Interestingly, no changes in the amplitude of the peaks in the swing angle's distribution were observed in this case. Since the difference in force constants for bonded potentials are minor, MARTINI 2.0 and 3.0 models differ primarily in the parametrization of LJ contributions and in the positioning of the center of the beads, located in the COM or the COG respectively. Aiming to explain why MARTINI 2.0 models did not exhibit any peak around 35° for the swing angle while MARTINI 3.0 models did, three new sets of simulations were performed: (i) changing the guest-host cross interactions of the MARTINI 2.0 models to be precisely those of the V MARTINI 3.0 model, (ii) changing the equilibrium relative position of the beads to the COG instead of the COM, and (iii) applying the two previous modifications simultaneously. None of these set-ups has shown any sign of improvement in the direction of reproducing the swing effect. This suggests that tuning the host-host LJ interactions, which were left unchanged in all the three simulation sets described above, may be also important in reproducing the swing effect. Notably, as discussed, MARTINI 3.0 models lie in the attractive region of the LJ potential. The influence of host-host LJ interactions, together with the influence of ε of guest-host LJ interactions, suggests that the key to designing a classical model able to reproduce the AP → HP transition may lie in successfully parametrizing both the host-host and host-guest LJ interactions at the same time.

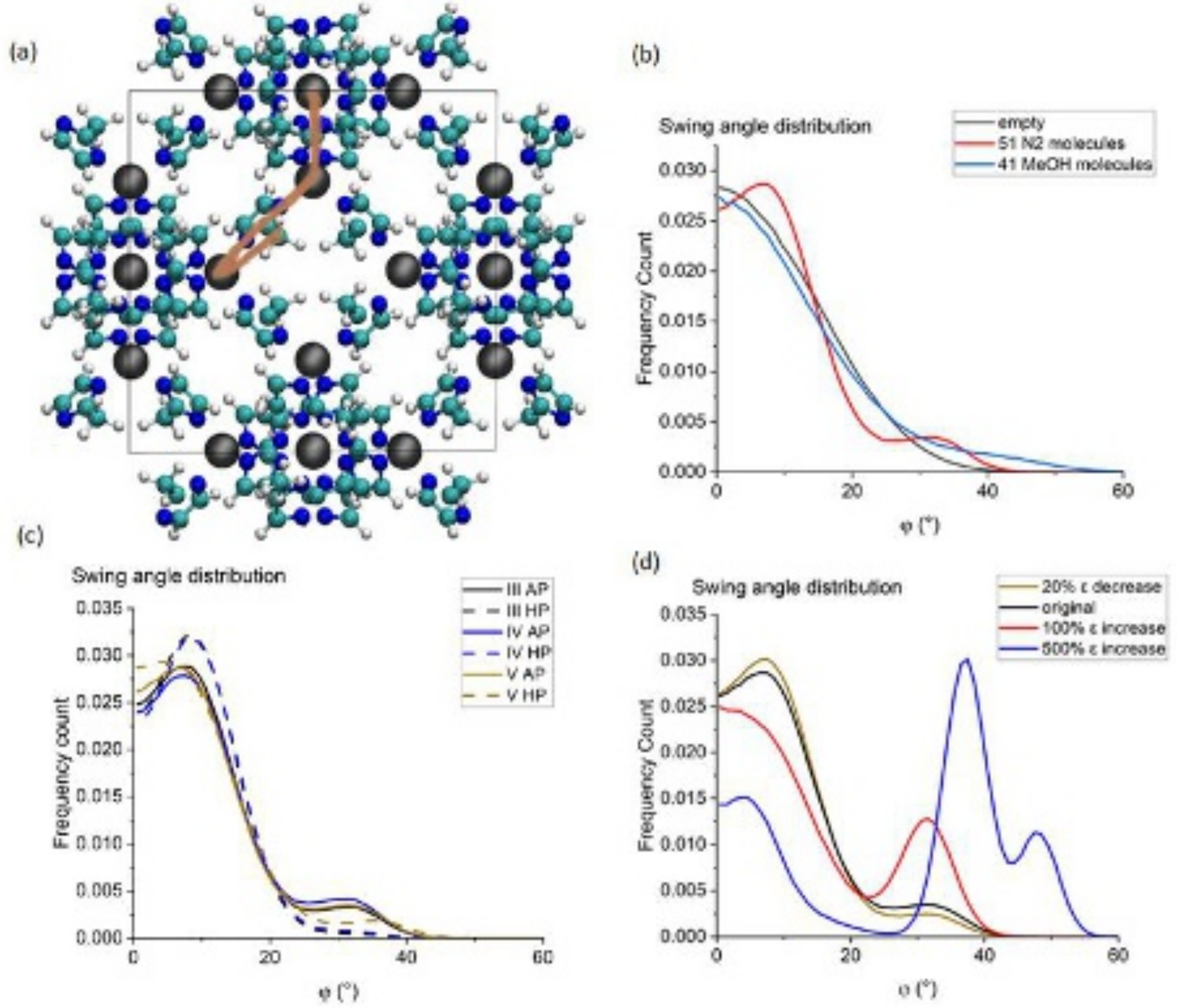

Figure 5. (a) Snapshot illustrating the dihedral angle, $\varphi$, (i.e. the swing angle), used to specify the position of the linkers with respect to the 6MR window. Color code: cyan (C), white (H), blue (N), gray (Zn). (b) Swing angle distributions for empty and loaded structures obtained from the mCM3V model. (c) Swing angle distributions obtained upon using as initial microstate for the MD simulations the AP (original approach) and HP structure using the mCM3III,, mCM3IV and mCM3V models. (d) Swing angle distributions for the loaded system (51 $N_2$ molecules) obtained upon using different values of ε of all the LJ cross-interactions changed by different percentages from the original value within the mCM3V model.

Finally, the normalized average pairwise interaction energies (LJ and Coulombic, when the latter exists) between guest and host, calculated considering superatoms of the MOF of a specific type, are presented in table 2 together with their standard deviation.

Ultimately, both mcM2 and mDM2 models tend to yield slightly more favorable guest-host interactions with the beads composing the linker than their M3 counterparts do. The value of $N_2$-(metal-containing-bead) (MOF beads 1) interactions remains quite stable for all mC and mD models, regardless of the MARTINI version and of the presence or absence of charges. Moreover, charged models have different effects in the two different MARTINI versions. In the M3 charged models, the linker exhibits less favorable interactions with the $N_2$ molecules compared to the non-charged models. Coulombic interactions within MARTINI 3.0 models (mCM3I, mCM3II and mDM3I) overall destabilize the system by increasing the interaction energies with the linker superatoms by about 50%.

| | | **Total pairwise bead 1-$N_2$ (kJ/mol)** | **Total pairwise bead 2-$N_2$ (kJ/mol)** | **Total pairwise bead 3-$N_2$ (kJ/mol)** |
|---|---|---|---|---|
| **mCM2** | **I** | -0.209 (SD = 0.014) | -0.123 (SD = 0.007) | -0.141 (SD = 0.007) |
| | **II** | -0.216 (SD = 0.013) | -0.112 (SD = 0.006) | -0.129 (SD = 0.007) |
| | **III** | -0.181 (SD = 0.013) | -0.112 (SD = 0.006) | -0.129 (SD = 0.007) |
| | **IV** | -0.217 (SD = 0.013) | -0.110 (SD = 0.006) | -0.129 (SD = 0.006) |
| | **V** | -0.182 (SD = 0.013) | -0.111 (SD = 0.006) | -0.129 (SD = 0.007) |
| **mCM3** | **I** | -0.239 (SD = 0.016) | -0.038 (SD = 0.008) | -0.057 (SD = 0.008) |
| | **II** | -0.230 (SD = 0.016) | -0.043 (SD = 0.009) | -0.057 (SD = 0.009) |
| | **III** | -0.241 (SD = 0.016) | -0.108 (SD = 0.009) | -0.093 (SD = 0.008) |
| | **IV** | -0.230 (SD = 0.016) | -0.108 (SD = 0.009) | -0.083 (SD = 0.008) |
| | **V** | -0.248 (SD = 0.016) | -0.107 (SD = 0.009) | -0.099 (SD = 0.008) |
| **mDM2** | **I** | -0.113 (SD = 0.018) | -0.160 (SD = 0.005) | -0.133 (SD = 0.010) |
| | **II** | -0.132 (SD = 0.017) | -0.173 (SD = 0.004) | -0.128 (SD = 0.009) |
| | **III** | -0.133 (SD = 0.016) | -0.143 (SD = 0.004) | -0.131 (SD = 0.009) |
| | **V** | -0.144 (SD = 0.018) | -0.151 (SD = 0.005) | -0.133 (SD = 0.009) |
| **mDM3** | **I** | -0.155 (SD = 0.016) | -0.052 (SD = 0.005) | -0.056 (SD = 0.009) |
| | **II** | -0.160 (SD = 0.014) | -0.089 (SD = 0.004) | -0.085 (SD = 0.008) |
| | **III** | -0.159 (SD = 0.014) | -0.089 (SD = 0.004) | -0.077 (SD = 0.008) |
| | **IV** | -0.170 (SD = 0.014) | -0.097 (SD = 0.004) | -0.092 (SD = 0.007) |

Table 2: Normalized average values of pairwise energy (LJ and, when exists, Coulombic) corresponding to the interaction between the $N_2$ molecules and superatoms of the MOF of a given type. The standard deviation is presented in-between parenthesis.

## d) Force field selection for the different mappings

The choice concerning which classical model to use in the scope of each mapping depends on the properties and features that are desired to be well reproduced, being a compromise sometimes necessary. As the associated standard deviation of physical and mechanical properties for models for a same mapping have similar order of magnitudes, model performance is assessed by comparing the mean values only.

For mapping A, mAM2 and mAM3II models are chosen as the best options for better reproducing the elastic tensor. As no relevant differentiation exists between them, the selection cannot be further refined. For mapping B, the best models differ depending on which properties are prioritized. mBM2 models predict better values of the elastic tensor whilst mBM3 excels in predicting the lattice parameter at (300K, 0 GPa). However, as the overprediction of the lattice parameter by MARTINI 2.0 models is in a much lower degree (on average by 1.54%) compared to the underprediction of elastic constants by MARTINI 3.0 models (on average by 28% for $C_{11}$, 63% for $C_{12}$ and 59% for $C_{44}$), the former are pointed as better options. Results on thermal expansion support electing model mBM2II as it is the most successful one in reproducing the reference value. For mapping C, similarly to mapping B, looking at the lattice parameter and mechanical properties would lead to choosing one of the MARTINI 2.0 models. Unlike for the previous mapping though, there are no significant differences in the thermal expansion coefficient for these models, thus dismissing a more refined selection. Lastly, for mapping D, as reproducing the structure is important for accurately reproducing many other properties and phenomena, all mDM3 models are chosen as better options. Models mDM3III or mDM3IV are preferred thanks to their better accuracy in reproducing the fitting targets in the present paper. As no model was able to capture the swing effect in the context of mappings C and D, no weight coming from it comes in the selection of the best model for each mapping.

## CONCLUSIONS

In this work MARTINI force fields are applied to modeling a MOF, namely ZIF-8, for the first time. The effects of the degree of coarsening, the choice of bead flavors and the MARTINI force field version are systematically investigated leading ultimately to several MARTINI models from versions 2.0 and 3.0 to be considered for four different mappings. The models were evaluated regarding their capability of reproducing RDFs, ADFs, lattice parameter, thermal expansion coefficient, elastic constants and guest-induced swing effect. Numerical instabilities were observed for models with both predominantly high attraction or high repulsion degrees coming from the LJ interactions relative to the superatom's spacing fixed by the bonded potentials, revealing the importance of having balanced LJ interactions.

Structure is well reproduced by all models under assessment except by MARTINI 2.0 models for mapping D, which are not able to reproduce proper peak alignment nor the overall profiles of many RDFs and ADFs. The lattice parameter at (300K, 1 atm), evaluated in the NPT ensemble, is overall well described by all the models. Overall, MARTINI 2.0 models tend to slightly overestimate the lattice parameter compared to their counterparts from version 3.0, and they drastically overestimate it in the case of mapping D. Values of $C_{11}$ and $C_{12}$ tend to be better reproduced by MARTINI 2.0 models, with MARTINI 3.0 models slightly underestimating them. No model particularly excels in reproducing the elastic tensor as a whole at (300K, 0 GPa). All models predict a decrease in $C_{44}$ value as the pressure at (300K, 0.2 GPa) from (300K, 0 GPa), which has been pointed out as a key feature for the amorphization of ZIF-8. However, none of them are able to picture the amorphization within the simulation time considered. Thermal expansion coefficients were more accurately reproduced by mDM2II and mDM2III. Bead flavor does not seem to impact on properties of the empty ZIF-8, although it does have an impact in host-guest interactions. An analysis of $N_2$ loaded ZIF-8 shows that M2 tends to yield more favorable host-guest interaction than M3. Presence of charges within M3 models tend to destabilize the host-guest interactions as its corresponding energy is larger. None of the MARTINI models considered for mappings C and D were able to capture the swing effect. A proper LJ parametrization for the guest-host and host-host interactions was suggested as a possible key feature in designing a CG model able to reproduce the effect, being further investigation on the topic needed.

We conclude that MARTINI force fields can be applied to studying ZIF-8 to a reasonable accuracy within the fitting targets considered in the present work. A global statement concerning whether MARTINI 2.0 or MARTINI 3.0 is better for modeling ZIF-8 in the CG level cannot be stated a priori, as there seems to be a tendency on force-fields coming from one version to better reproduce certain properties and features of the system but not others. Given the importance of properly predicting the structure of the system, using a MARTINI 3.0 model is generally advised when the mapping of choice is less coarsened. This is no surprise given that this MARTINI version has been more carefully parameterized to suit less coarsened systems. We hope that this work will act as a springboard to use MARTINI force fields within the MOF community to investigate mesoscale phenomena by computer simulations.

## SUPPLEMENTARY MATERIAL

The supplementary material is free of charge at …. It contains complete RDFs, ADFs and BDFs for all the optimized MARTINI models used to model ZIF-8 as well as a table precisely informing their parametrization. It also contains a table that allows to see more readily the value of $R_{min}$ for each pair of superatom types prescribed by the LJ potential of each model in contrast with the actual distance between this given pair, observed in the equilibrium structure. Finally, it contains histograms of swing angle values for the models of mapping C and D for the empty framework as well as the loaded conditions investigated.

## ACKNOWLEDGEMENTS

The authors thank the École Doctoral Sciences Chimiques Balard for funding this work. This work was granted access to the HPC resources of CINES under the allocation A0110911989 made by GENCI.

## DATA AVAILABILITY STATEMENT

The data that support this article are available within the article and the supplemental material. Some useful additions (input files, codes) are also provided in https://github.com/rosemino/MARTINI_ZIF-8.

## Author Contribution

C. A. run all atomistic and CG simulations. R. S. designed the study. G. M. and R. S. supervised the work. C. A. and R. S. wrote the manuscript and created the figures. All authors analyzed and discussed the results and read and approved the manuscript.

## Conflict of Interests

The authors have no conflicts to disclose.

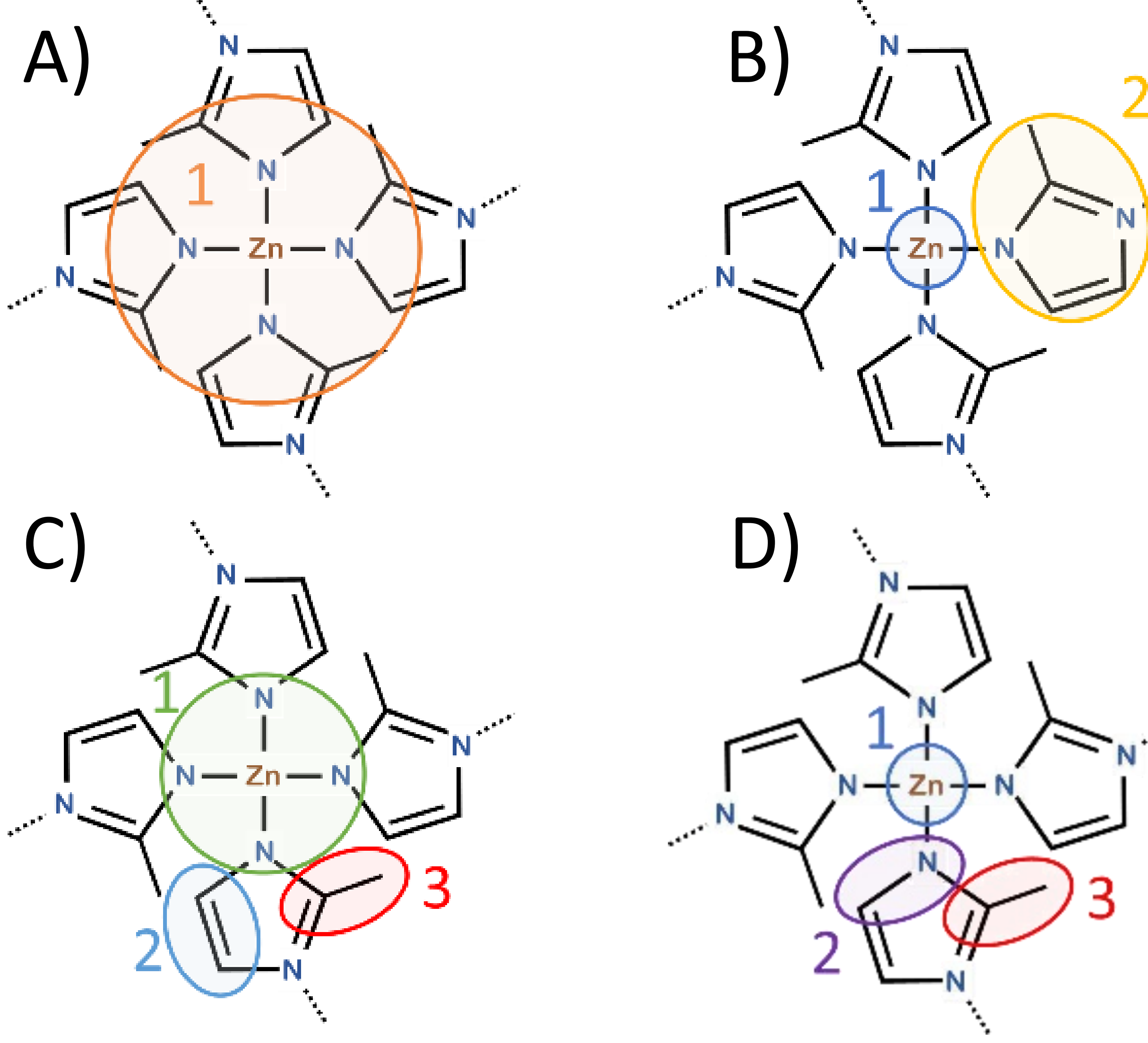

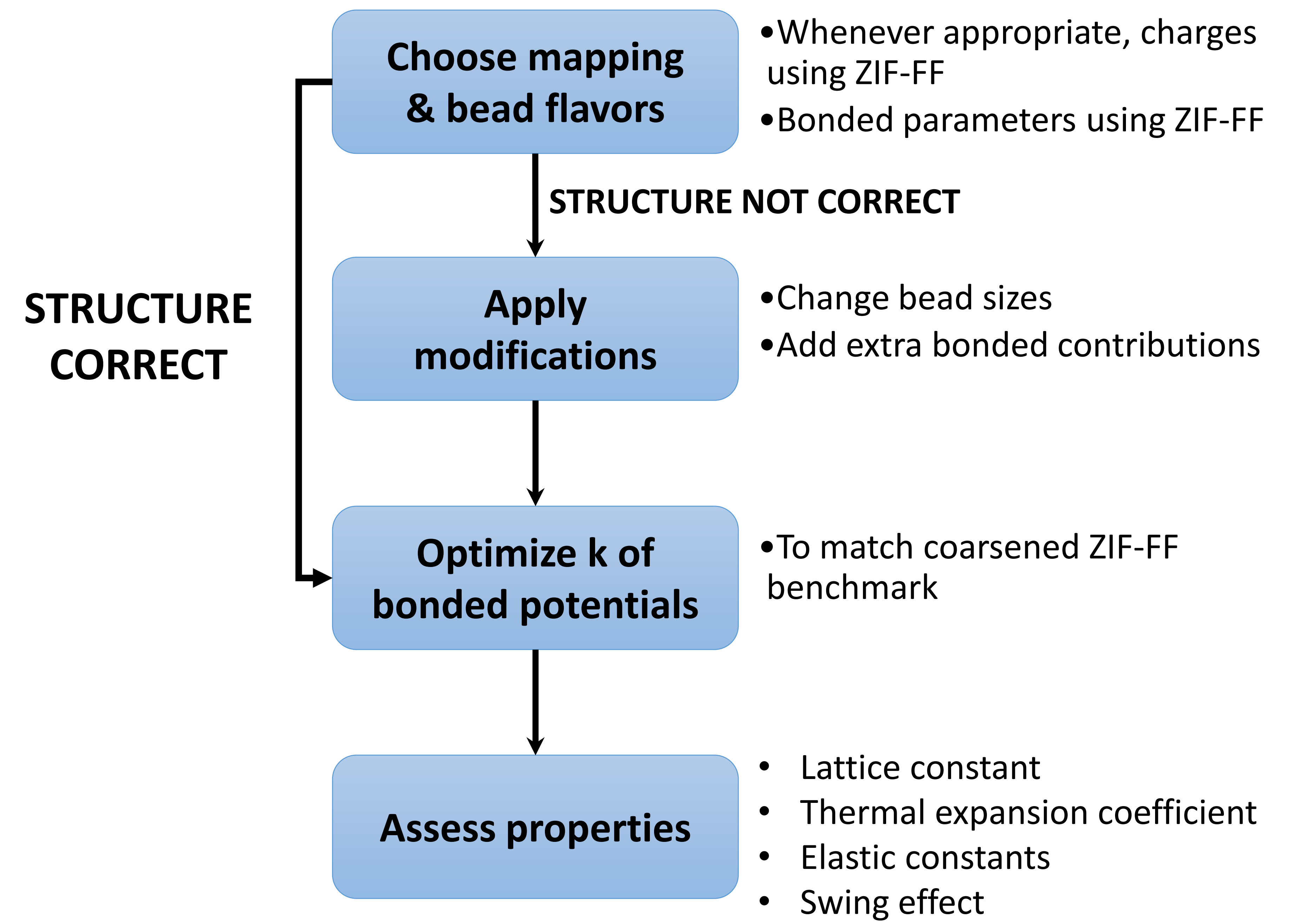

Choose mapping & bead flavors
•Whenever appropriate, charges using ZIF-FF
•Bonded parameters using ZIF-FF
STRUCTURE NOT CORRECT
STRUCTURE CORRECT
Apply modifications
•Change bead sizes
•Add extra bonded contributions
Optimize k of bonded potentials
•To match coarsened ZIF-FF benchmark
Assess properties
• Lattice constant
• Thermal expansion coefficient
• Elastic constants
• Swing effect

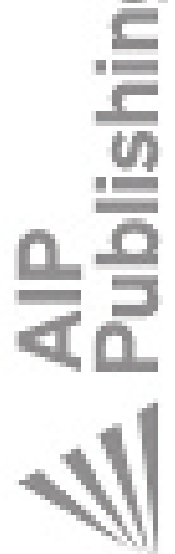

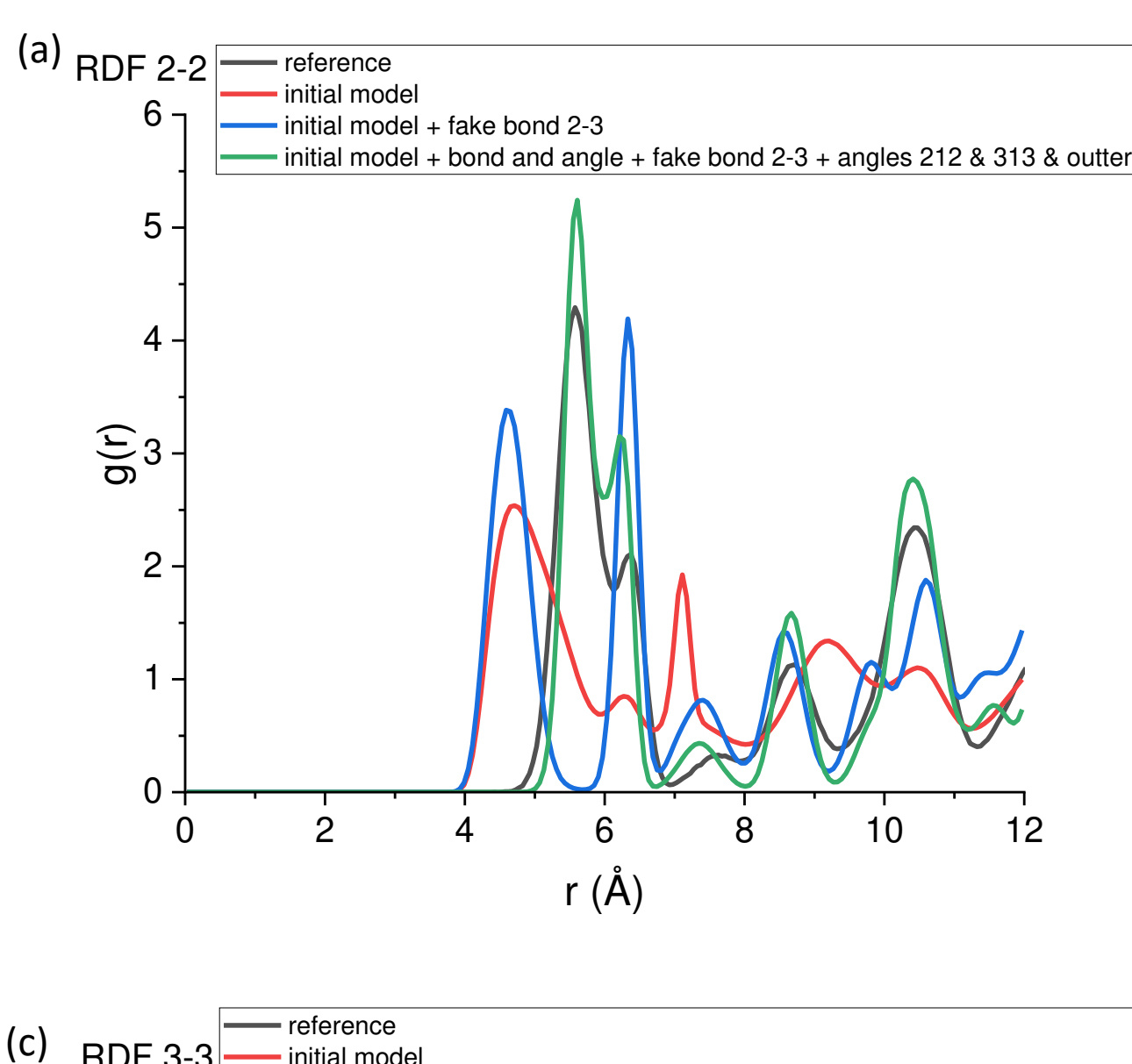

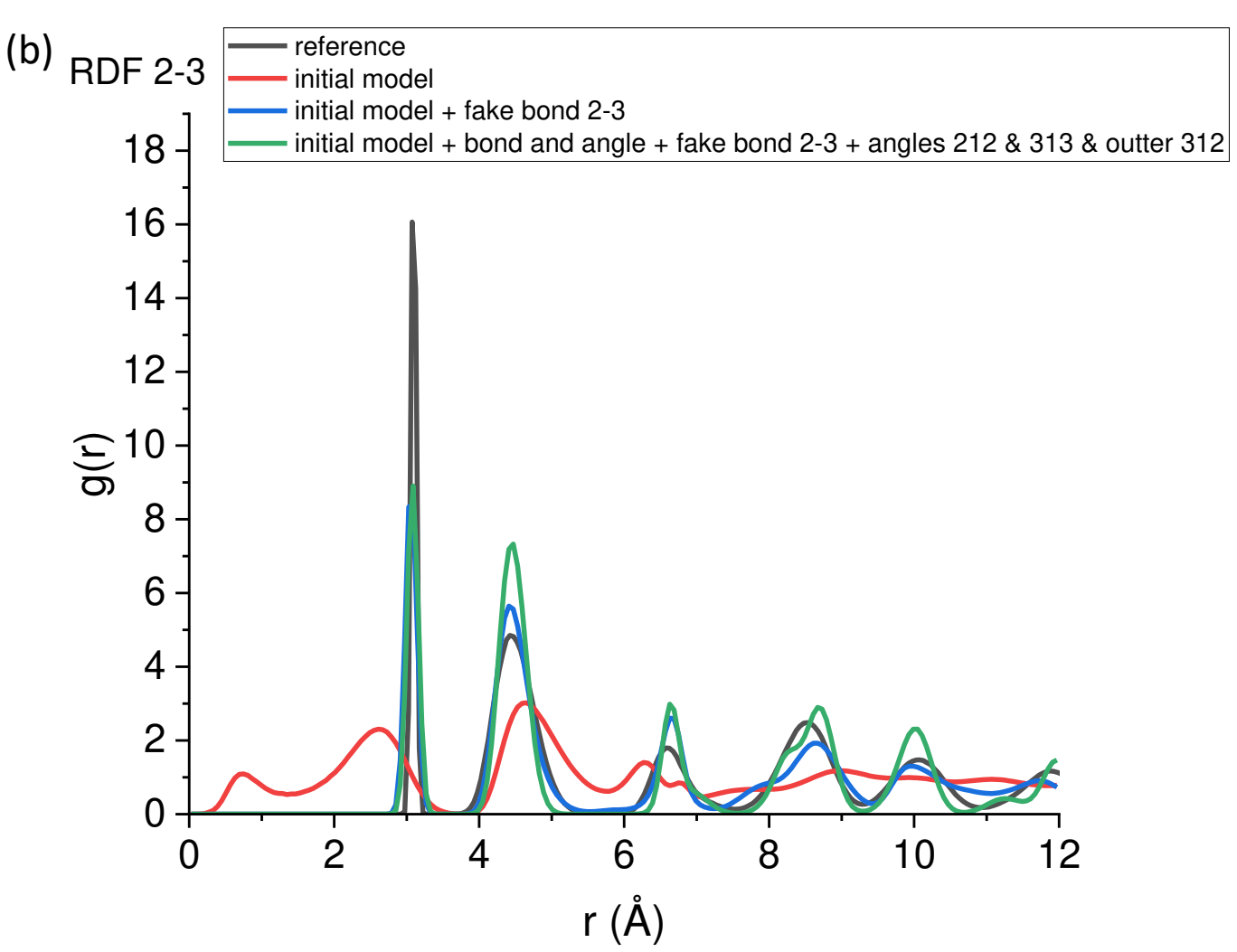

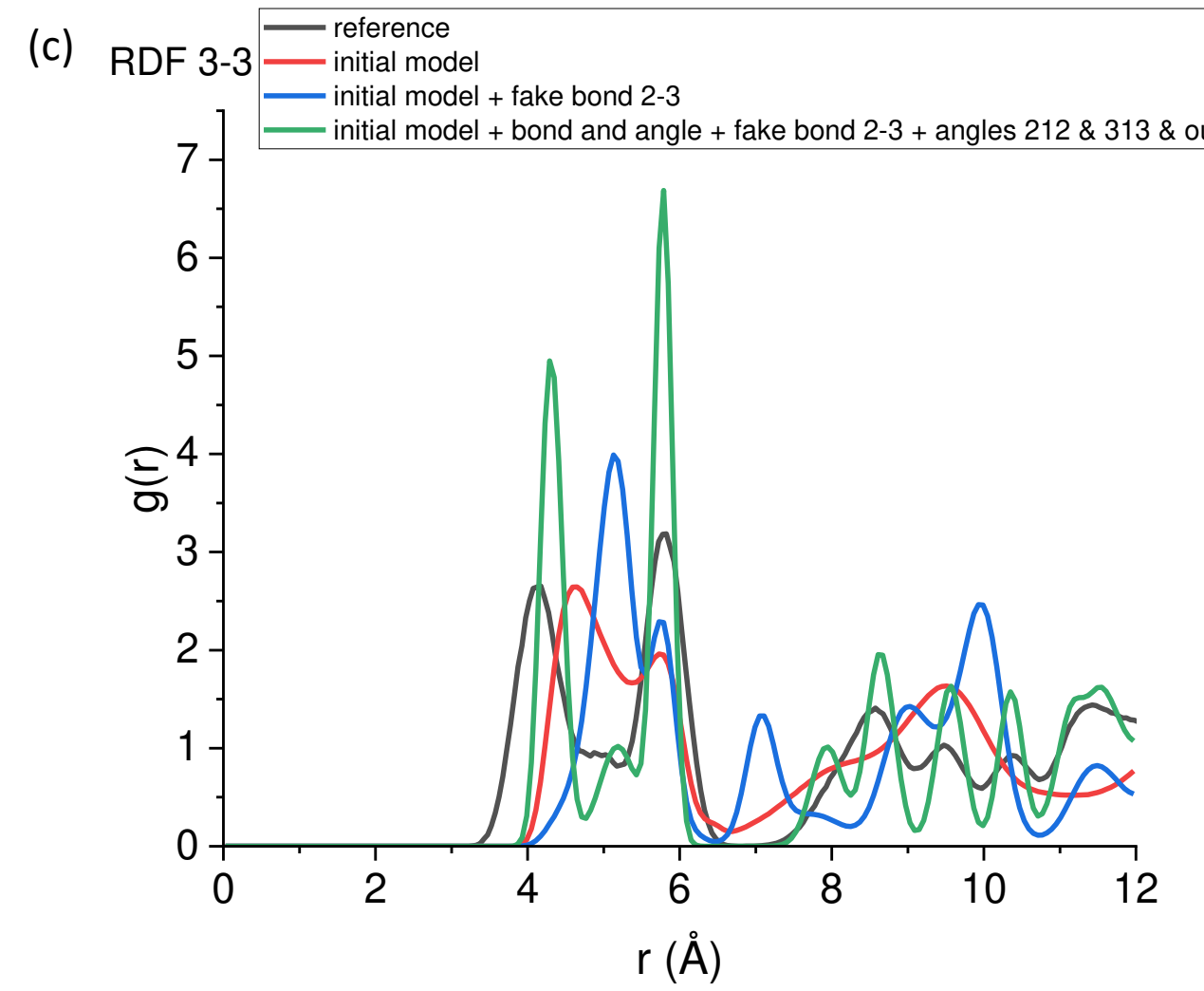

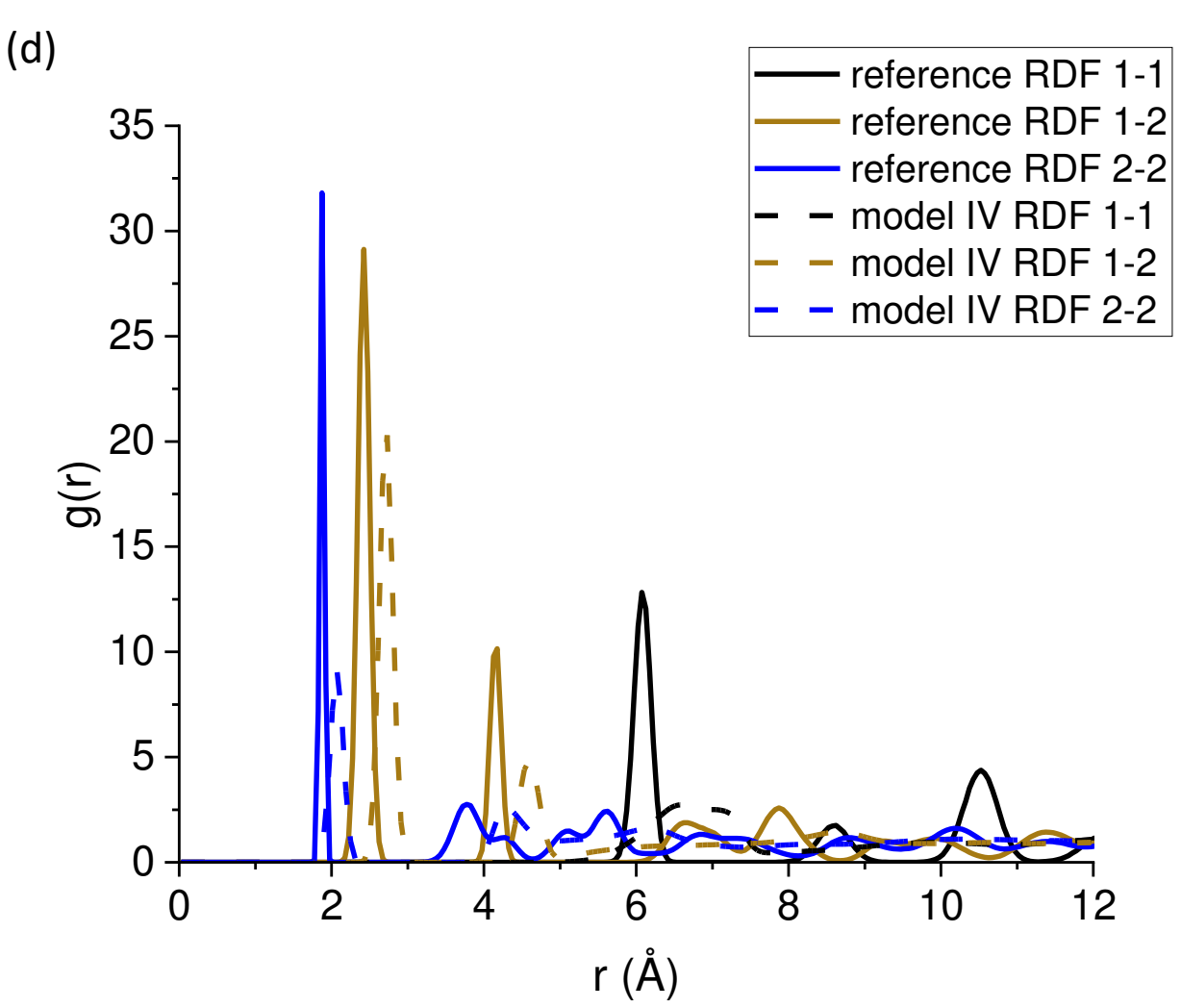

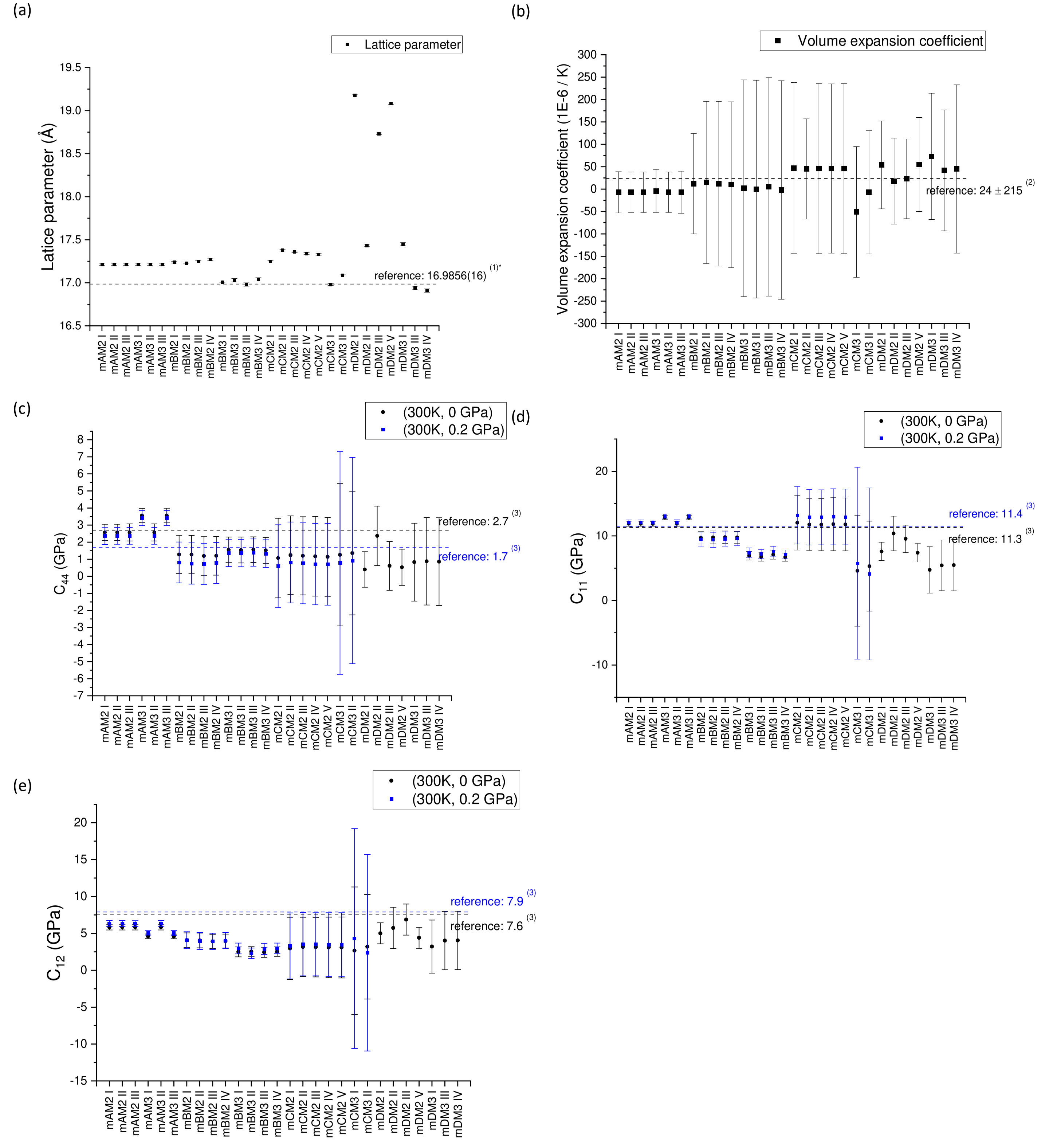

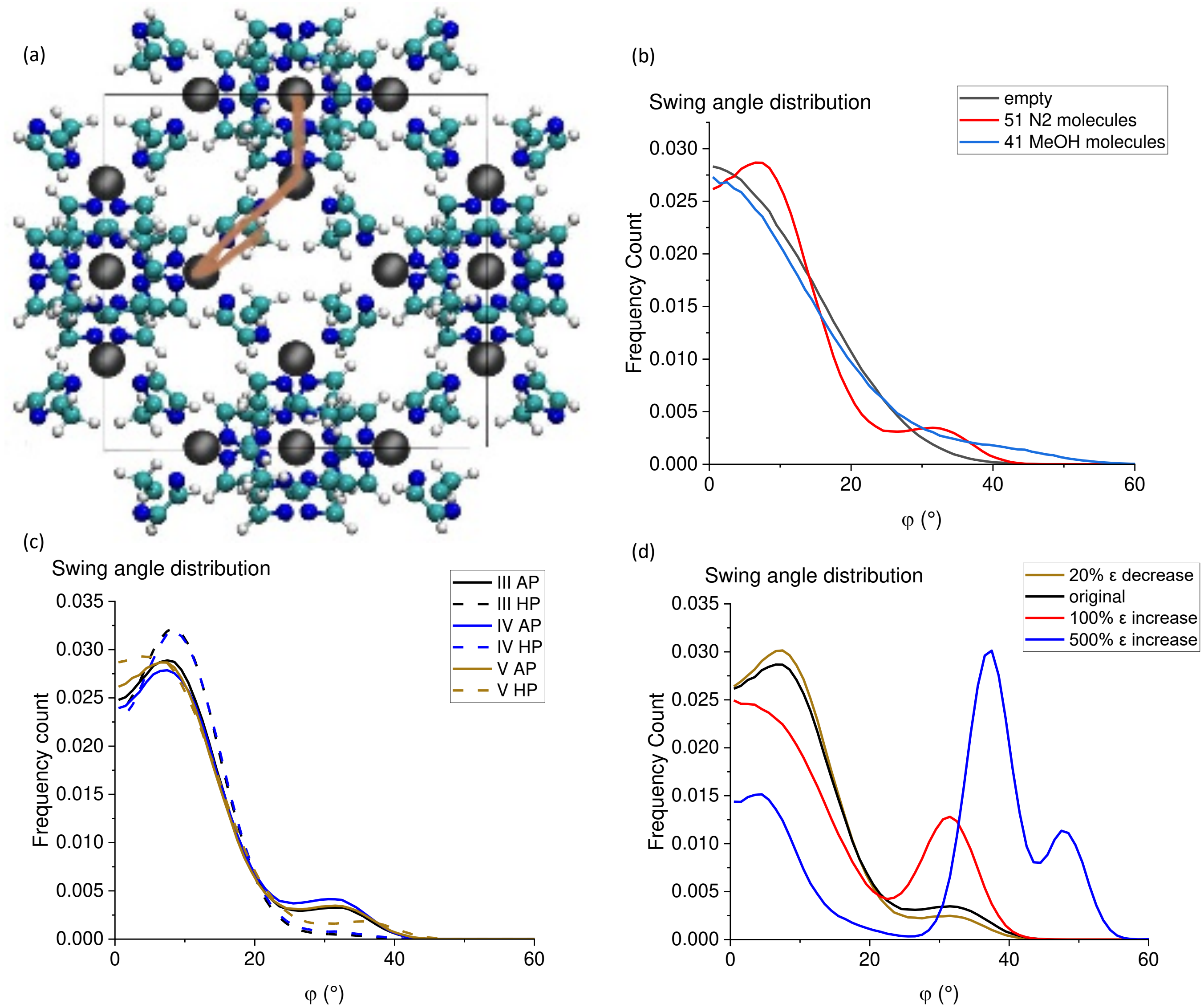

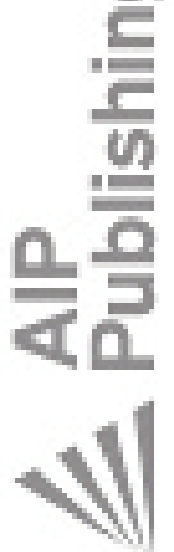